\documentclass[aip, jmp, amsmath,amssymb,preprint,nofootinbib%
]{revtex4-1}
\usepackage[latin9]{inputenc}
\usepackage{amsmath}
\usepackage{amssymb}
\usepackage{esint}
\usepackage{graphicx}
\usepackage{verbatim}

\providecommand{\tabularnewline}{\\}

\makeatother

\begin{document}
\title{Magnetic properties of a long, thin-walled ferromagnetic nanotube}
\author{Chen Sun}
\affiliation{Department of Physics, Texas A\&M University, College Station, Texas 77843-4242, USA}
\author{Valery L. Pokrovsky}
\affiliation{Department of Physics, Texas A\&M University, College Station, Texas 77843-4242, USA}
\affiliation{Landau Institute for Theoretical Physics, Chernogolovka, Moscow District, 142432, Russia}

\date{\today}

\begin{abstract}
We consider magnetic properties of a long, thin-walled ferromagnetic
nanotube. We assume that the tube consists of isotropic homogeneous
magnet whose spins interact via the exchange energy, the dipole-dipole
interaction energy, and also interact with an external field via Zeeman
energy. Possible stable states are the parallel state with the magnetization
along the axis of the tube, and the vortex state with the magnetization
along azimuthal direction. For a given material, which of them has
lower energy depends on the value $\gamma=R^{2}d/(L\lambda_{x}^2)$,
where $R$ is the radius of the tube, $d$ is its thickness, $L$
is its length and $\lambda_{x}$ is an intrinsic scale of length characterizing
the ration of exchange and dipolar interaction. At $\gamma<1$ the
parallel state wins, otherwise the vortex state is stable. A domain
wall in the middle of the tube is always energy unfavorable, but it
can exist as a metastable structure. Near the ends of a tube magnetized
parallel to the axis a half-domain structure transforming gradually
the parallel magnetization to a vortex just at the edge of the tube
is energy favorable. We also consider the equilibrium magnetization
textures in an external magnetic field either parallel or perpendicular
to the tube. Finally, magnetic fields produced by a nanotube and an
array of tubes is analyzed. 

\end{abstract}
\maketitle
\section{INTRODUCTION}

Magnetic nanomaterials play an important role in applications as elements
of memory and magnetic sensors and switches as it was demonstrated
by Nobel prize 2007 to Fert and Gr\"{u}nberg for their invention of antiferromagnetic
spin valve. The task of further miniaturization of magnetic devices
and creation of configurations providing a controllable magnetic field
is extremely important for nanophysics and technology. Experimenters
and technologists have already created nanomagnets in different shapes
-- disks \cite{disk}, rings \cite{ring}, wires \cite{nanowire},
etc. Among these new nanomaterials the nanotubes, as compared to solid
wires, have inner voids that reduce the density of materials and makes
them easier to float in solutions, a desirable property in biotechnology.
\cite{equilibrium states} The inner hollow itself can be used for
capturing large biomolecules. \cite{chemistry} Besides, as magnetic
materials, they are free of vortex cores, which makes the vortex state
more stable than that of nanowires. This makes nanotubes more suitable
as candidates for elements of memory for computers and as a tool for
creation of superconductors with high critical fields. Several methods
have been used to synthesize nanotubes: electrodeposition \cite{Nielsch 3,electrodeposition},
atomic layer deposition \cite{ALD}, hydrogen reduction \cite{hydrogen reduction}.
Ferromagnetic materials used for formation of nanotubes include Ni,
\cite{ALD}, Co \cite{Nielsch 3,ALD}, FePt \cite{hydrogen reduction},
$\mathrm{Fe_{3}O_{4}}$ \cite{hydrogen reduction}.

Together with the experimental progress, theoretical calculations
and numerical simulations for nanotubes were performed extensively,
dealing with the stable states \cite{phase diagrams,equilibrium states},
switching behavior \cite{reversal modes}, hysteresis loop \cite{equilibrium states,angular dependence}
and properties of domain walls (both static \cite{reversal modes}
or dynamic \cite{dynamic DW}). In Ref. \cite{phase diagrams} the
authors calculated numerically and partly analytically energy of the
parallel state and the vortex state as function of the dimensions
of the tube and material constants. Phase diagrams were drawn in terms
of linear dimensions of the tube. In Ref. \cite{equilibrium states}
the authors have shown that the parallel magnetization turns into
a vortex-like one at the edge of the tube.

The purpose of our work is to give an analytical description of the
magnetic tubes (MT) properties employing small parameters characterizing
their geometry: the ratios $d/R$ and $R/L$. In the experimentally
realized MT the first ratio was in the range of $10^{-3}$ and the
second one varied between $10^{-2}$ and $10^{-1}$. The analytical
approach allows as to construct the complete phase diagram of the
MT in the space of geometric parameters and external magnetic field.
We establish analytical criteria for the appearance and disappearance
of different topological magnetic configuration, topological defects
and field-induced magnetic textures. We also calculate the magnetic
field produced by the tubes.

\section{THE MODEL}

We take into account the magnetic interactions of two kinds: the exchange
interaction and the dipolar interaction. The total energy of a MT
is:
\begin{eqnarray}
E&=&E_{exch}+E_{dip}\nonumber\\
&=&-J\sum_{<\mathbf{x},\mathbf{x'}>}\mathbf{S_{x}}\cdot\mathbf{S_{x'}}+\frac{\mu_{0}}{4\pi}(g\mu_{B})^{2}\sum_{\mathbf{x},\mathbf{x'},\mathbf{x}\ne\mathbf{x'}}\frac{\mathbf{S_{x}}\cdot\mathbf{S_{x'}}-3(\mathbf{S_{x}}\cdot\mathbf{\hat{r}})(\mathbf{S_{x'}}\cdot\mathbf{\hat{r}})}{r^{3}}.\label{}
\end{eqnarray}
Here $\mathbf{S_{x}}$ is the spin vector at position $\mathbf{x}$,
$J$ is the exchange constant, $\mathbf{\hat{r}}$ is the unit vector
from position $\mathbf{x}$ to $\mathbf{x'}$, $<\mathbf{x},\mathbf{x'}>$
means summation over all nearest pairs, and $r=|\mathbf{r}|=|\mathbf{x}-\mathbf{x'}|$.
\cite{griffiths} Further we use the International System of units.
Another often considered contribution to the total energy, the crystal
anisotropy is not included here. This is appropriate 
when the material is a polycrystal with large number of randomly orientated
grains like permalloy. Experimenters \cite{Nielsch 3} indicate that
the size of a single-crystal grain in their nanotubes is about 1nm.
We do not know how strong is the exchange interaction between the
grains. In our calculations we assume that it is the same as in the
bulk single crystal.

We accept an approximation of classical continuous field $\mathbf{m(x)}$
for the magnetic order parameter with the constraint $\mathbf{m}^{2}(\mathbf{x})=1$.
The magnetization at the point $\mathbf{x}$ of the space is equal
to $M_{0}\mathbf{m}(\mathbf{x})$. The saturation magnetization $M_{0}$
is assumed to be dependent on temperature, but independent of the
point $\mathbf{x}$ of the space. In this approximation \cite{aharoni}
the exchange energy reads:
\begin{equation}
E_{exch}=A\int d^{3}x(\mathbf{\nabla m(x)})^{2},\label{exchange}
\end{equation}
where $A=\frac{1}{6}Jns^{2}Za^{2}$, $n$ is the density of magnetic
atoms, $s$ is the magnitude of their spin, $(\mathbf{\nabla m(x)})^{2}=(\nabla m_{x}(\mathbf{x}))^{2}+(\nabla m_{y}(\mathbf{x}))^{2}+(\nabla m_{z}(\mathbf{x}))^{2}$,
$n$ is the number of magnetic atoms per unit volume, $a$ is the
distance between two nearest atoms, $Z$ is the coordination number.

There are several equivalent expressions for the dipolar energy: 

\begin{subequations}
\begin{equation}\label{dip1}
E_{dip}=\frac{1}{2}\frac{\mu_{0}}{4\pi}M_{0}^{2}\int d^{3}xd^{3}x'\frac{\mathbf{m(x)}\cdot\mathbf{m(x')}-3(\mathbf{m(x)}\cdot\mathbf{\hat{r}})(\mathbf{m(x')}\cdot\mathbf{\hat{r}})}{r^{3}},
\end{equation}
\begin{equation}\label{dip2}
E_{dip}=\frac{1}{2}\frac{\mu_{0}}{4\pi}M_{0}^{2}\int d^{3}xd^{3}x'(\mathbf{m(x)}\cdot\mathbf{\nabla_{x}})(\mathbf{m(x')}\cdot\mathbf{\nabla_{x'}})\frac{1}{r},
\end{equation}
\begin{eqnarray}\label{dip3}
E_{dip}&=&\frac{1}{2}\frac{\mu_{0}}{4\pi}M_{0}^{2}\left[\int dAdA'\frac{\sigma_{M}(\mathbf{x})\sigma_{M}(\mathbf{x'})}{r}+2\int dAd^{3}x'\frac{\sigma_{M}(\mathbf{x})\rho_{M}(\mathbf{x'})}{r}\right.\nonumber\\
&&\left.+\int d^{3}xd^{3}x'\frac{\rho_{M}(\mathbf{x})\rho_{M}(\mathbf{x'})}{r}\right].
\end{eqnarray}
\end{subequations}

The integration denoted by $\int dA,\int dA'$ proceeds over the surfaces
of the magnet, the integration denoted as $\int d^{3}x$ goes over
its volume. The value $\sigma_{M}(\mathbf{x})=\mathbf{m(x)}\cdot\mathbf{n}$
is the ``surface magnetic charge density'' and $\rho_{M}(\mathbf{x})=-\nabla_{x}\cdot\mathbf{m(x)}$
is the ``volume magnetic charge density''. Eq. \eqref{dip3} is a form analogous to the energy
of electric charges interacting via Coulomb forces. This analogy allows
to use results well-known in electrostatics. An important consequence
of this analogy is that the dipolar energy is non-negative, since
the electrostatic energy is equal to the integral of the square of
the electric field.
Eq. \eqref{dip3} provides a clear electrostatic visualization of the
dipolar interaction. Eq. \eqref{dip2}
may occur more convenient for specific calculations. A system of magnetic
charges is always neutral. 

\section{STABLE STATES}

\begin{figure}
  \centering
  \includegraphics[width=0.6\textwidth]{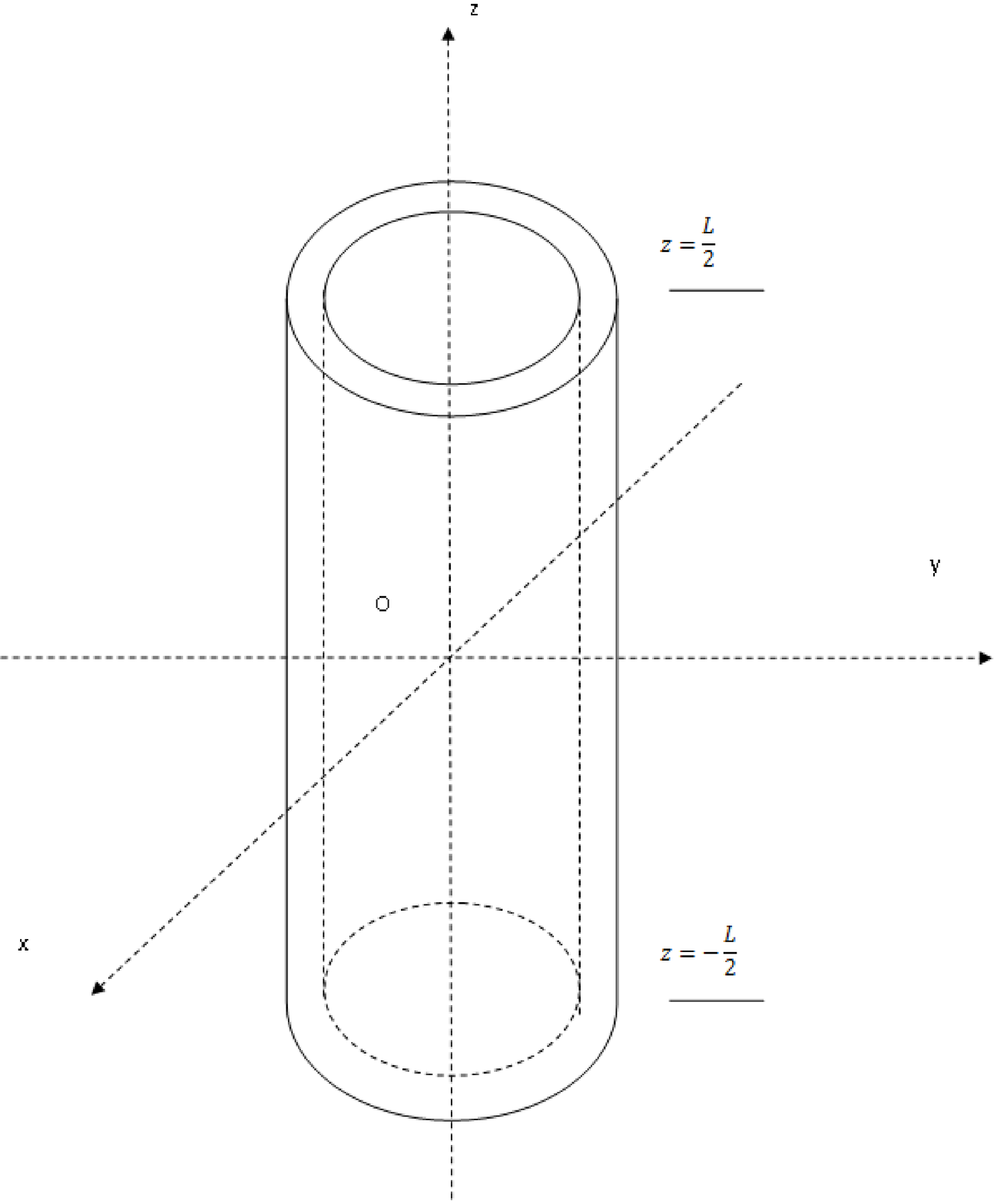}\\
  \caption{Geometry of a tube, with two edges located at $z=L/2$ and $z=-L/2$.}\label{1}
\end{figure}

We consider a cylindrical tube located between $z=-\frac{L}{2}$ and
$z=\frac{L}{2}$, as shown in Fig. \ref{1}, with the radius $R$, thickness
$d$, and length $L$. We assume $d\ll R\ll L$. This research was
initially stimulated by a new material fabricated experimentally by Dr. Wenhao Wu and his group
at Texas A\&M University: an array of nickel nanotubes in alumina
with dimensions approximately $R=150nm$, $d=30nm$ and $L=60\mu m$.
For these nanotubes the condition $R\ll L$ is well satisfied, while
the condition $d\ll R$ is relatively not so well satisfied. Besides,
$R^{2}\ll dL$ is also well satisfied. In earlier experiments {[}1-9{]}
all three strong inequalities were satisfied.

\begin{figure}
  \centering
  \includegraphics[width=0.6\textwidth]{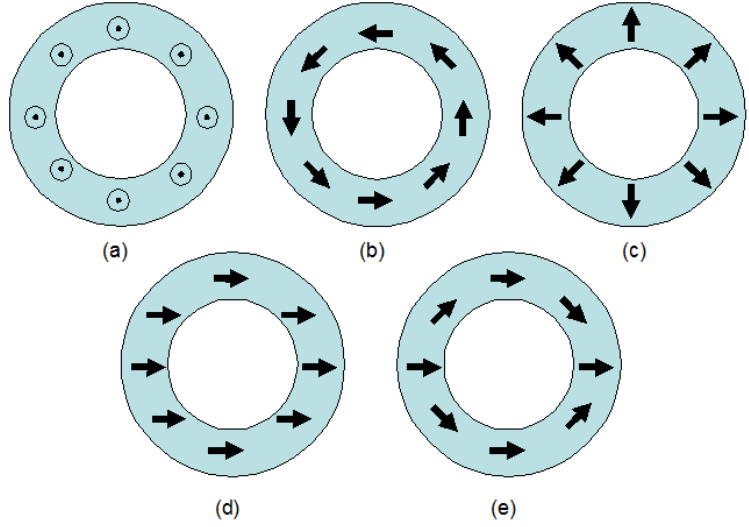}\\
  \caption{Pictorial characteristic of different magnetization distributions: (a) parallel; (b) vortex; (c) radial; (d) transverse; (e) onion.}\label{2}
\end{figure}

Natural candidates 
to the state with the lowest
energy are 
the most symmetric magnetic configurations: the parallel state:
$\mathbf{m(x)}=\hat{z}$, and the vortex state: $\mathbf{m(x)}=\hat{\phi}$.
We denote azimuthal angle as $\phi$; the symbol $\mathbf{m(x)}=\hat{\phi}$
denotes the unit vector in azimuthal direction. 
Each of these states is two-fold degenerate due to time reversal invariance.
We also consider two other, less symmetric states: the transverse
state: $\mathbf{m(x)}=\hat{x}$, and the so-called onion state. According
to Ref. \cite{bland review}, the onion state 
(see Fig. \ref{2}), becomes stable 
in ferromagnetic
rings in some range of parameters. Here we consider its analogue for
a tube.
These two kinds of states occur to be stable magnetic configuration in the transverse
magnetic field. 

In the parallel state, $\mathbf{\nabla m(x)}=0$, $\rho_{M}(\mathbf{x})=0$,
$\sigma_{M}(\mathbf{x})=\pm1$ for $z=\pm\frac{L}{2}$, respectively,
and $\sigma_{M}(\mathbf{x})=0$ elsewhere. The exchange energy is
zero. The dipolar energy consists of three parts: the self-energies
of the two edges and the energy of interaction between them two. Since
$R\ll L$, the latter term is much smaller than the former two and
further we neglect it. The distance $r$ between two points with the
cylindrical coordinates $(\rho,\phi,z)$ and $(\rho^{\prime},\phi^{\prime},z^{\prime})$
belonging to a MT satisfying the inequality $d\ll R$ reads: $r\approx\sqrt{(\rho-\rho')^{2}+2R^{2}[1-\cos(\phi-\phi')]+(z-z')^{2}}$.
Each self-energy term after integration over $\phi$ and $\phi'$
is reduced to an integral of 
the complete elliptic integral
of the first kind $K(k)$ where $k=\sqrt{\frac{4\rho\rho'}{(\rho+\rho')^{2}}}$.
The condition $d\ll R$ allows us to use the approximation $K(k)\thickapprox2\log2-\frac{1}{2}\log(1-k^{2})$
since $k$ is close to 1. In this approximation the integration over
$\rho$ and $\rho'$ is straightforward leading to the result for
the self-energy of each edge: $E(\textrm{edge})=\frac{1}{2}\mu_{0}M_{0}^{2}Rd^{2}(\log\frac{8R}{d}+\frac{3}{2})$.
Thus, in the limit of long thin MT the total energy of the parallel
state is: $E_{P}=\mu_{0}M_{0}^{2}Rd^{2}(\log\frac{8R}{d}+\frac{3}{2})$.
It does not depend on the tube length $L$.

Different magnetic configurations and results of similar calculations
of exchange and dipolar energy for them are summarized in Table 1.

\bigskip{}
\makebox[1\textwidth]{%
Table 1. Magnetic configurations and energies.%
}

\smallskip{}
\makebox[1\textwidth]{%
\begin{tabular}{|c|c|c|c|}
\hline
state  & magnetization vector  & exchange energy  & dipolar energy\tabularnewline
\hline
parallel(P)  & $\hat{z}$  & 0  & $\mu_{0}M_{0}^{2}Rd^{2}(\log\frac{8R}{d}+\frac{3}{2})$\tabularnewline
\hline
vortex(V)  & $\hat{\phi}$  & $2\pi A\frac{Ld}{R}$  & 0\tabularnewline
\hline
radial(R)  & $\hat{\rho}$  & $2\pi A\frac{Ld}{R}$  & $\pi\mu_{0}M_{0}^{2}RdL$ \tabularnewline
\hline
transverse(T)  & $\hat{x}$  & 0  & $\frac{\pi}{2}\mu_{0}M_{0}^{2}RdL$ \tabularnewline
\hline
onion(O)  & $\hat{x}\cos(a\sin2\phi)-\hat{y}\sin(a\sin2\phi)$  & $4\pi A\frac{Ld}{R}a^{2}$  & $\frac{\pi}{2}\mu_{0}M_{0}^{2}RdL[1-J_{1}(2a)]$ \tabularnewline
\hline
\end{tabular}%
} \bigskip{}

While other configurations are trivial, some comments on the onion
configuration are necessary. We seek for a variational distribution
of magnetization that satisfies following requirements: magnetization
must be parallel to magnetic field (in the direction $\hat{x}$) at
$\phi=0,\pm\pi/2,\pi$ and it does not depend on the coordinate $r$
in a narrow ring. A simplest vector field satisfying these requirements
has a form $\mathbf{m}(\phi)=\hat{x}\cos\theta(\phi)+\hat{y}\sin\theta(\phi)$
with $\theta(\phi)=-a\sin(2\phi)$, where $a$ is a variational parameter.
Its value should minimize the total energy. ($a=0$ corresponds to
the transverse state.) As it is seen from the last line of the Table
1, the minimum energy corresponds to a largest maximum of the Bessel
function $J_{1}(2a)$=0.583. It corresponds to $a=0.92$, and $E_{O}=\frac{\pi}{2}\mu_{0}M_{0}^{2}RdL*0.418$.
It is smaller than $E_{T}$. Since the exchange energy is quadratic
in $a$, the minimum of the total onion configuration energy shifts
to a value of $a$ between $0$ and $0.92$. It is easy to check that
the difference of the total onion energy and transverse energy at
$a=0$ is zero, whereas the derivative of this difference over $a$
is negative. Therefore, its value in the minimum of the onion energy
is negative. In other words, the onion is energy preferable to the
transverse configuration. If $R\gg\lambda_{x}$, where $\lambda_{x}=\sqrt{A/(\mu_{0}M_{0}^{2})}$
is the so-called exchange length, the exchange interaction of the
onion configuration is much smaller than the dipolar term, so $a$
is very close to $0.92$. A typical value of $\lambda_{x}$ is several
nanometers. \cite{exchange length} 

\begin{figure}
  \centering
  \includegraphics[width=0.6\textwidth]{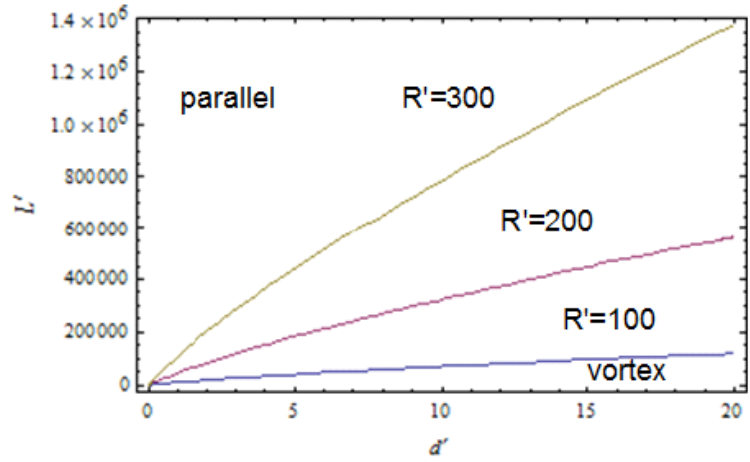}\\
  \caption{Phase boundaries between the parallel state and the vortex state. The region above a transition line corresponds to the parallel state. The three transition lines from bottom to top corresponds to $R'=100, 200$ and 300 respectively.}\label{3}
\end{figure}

The parallel and transverse states have zero exchange energy; the
vortex state has 
zero dipolar energy. In the absence of magnetic
field the hierarchy of energy scales is as follows: $E_{T}>E_{O}\gg E_{P}$,
so the transverse and the onion states are not stable. Only the parallel
or the vortex states are stable.\footnote{At small radius $R$, it may happen that $E_{V}>E_{T}$. The condition of such stabilization of the transverse configuration
is $\frac{R}{\lambda_{x}}<2$.} Which of these two states is more favorable depends on the ratio:
\begin{equation}
\gamma=\frac{E_{P}}{E_{V}}=\frac{R^{2}d}{2\pi\lambda_{x}^2L}\left(\log\frac{8R}{d}+\frac{3}{2}\right).\label{}
\end{equation}
At $\gamma<1$ the parallel states wins, in opposite case the vortex
state wins.
The logarithmic factor varies comparatively slowly,
and $\gamma$ depends mainly on the combination $\frac{R^{2}d}{L}$.
We see that large $L$ favors the parallel magnetization, while large
$R$ and $d$ 
favor the vortex magnetization. 
It is convenient to introduce dimensionless lengths $R'=\frac{R}{\lambda_{x}}$,
$d'=\frac{d}{\lambda_{x}}$, $L'=\frac{L}{\lambda_{x}}$. Then $\gamma$
can be expressed as
\begin{equation}
\gamma=\frac{R'^{2}d'}{2\pi L'}\left(\log\frac{8R'}{d'}+\frac{3}{2}\right).\label{}
\end{equation}
Equation $\gamma=1$ determines the transition line between the two
states. In Fig. \ref{3} three transition lines in the $(L^{\prime},d^{\prime})$
plane at fixed values of $R^{\prime}$ are depicted.

The external field can stabilize the transverse or the onion state.
This problem will be analyzed in section \ref{transverse external field}.

\section{DOMAIN WALLS}

\label{DW} 
Let us consider a tube with one domain wall (DW) far from the edges.
Calculations in this section are performed in the limit $L\rightarrow\infty$.

\begin{figure}
  \centering
  \includegraphics[width=0.6\textwidth]{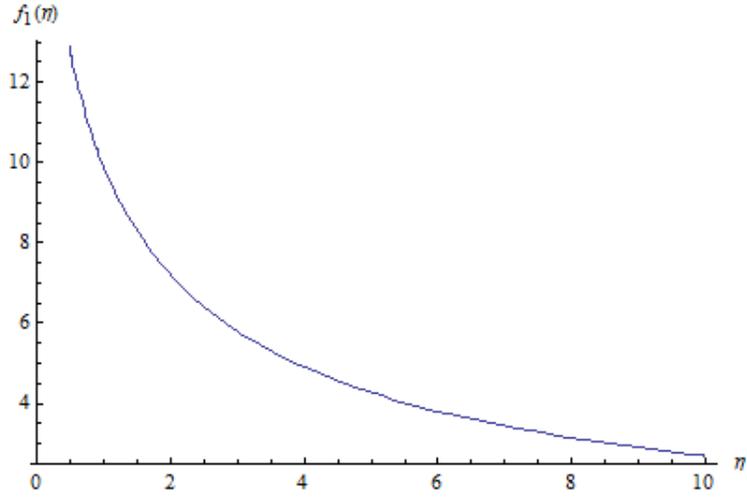}\\
  (a)\\
  \includegraphics[width=0.6\textwidth]{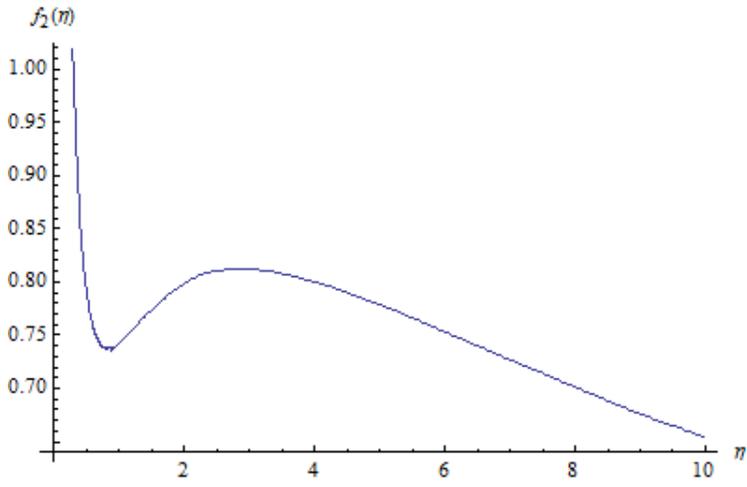}\\
  (b)\\
  \includegraphics[width=0.6\textwidth]{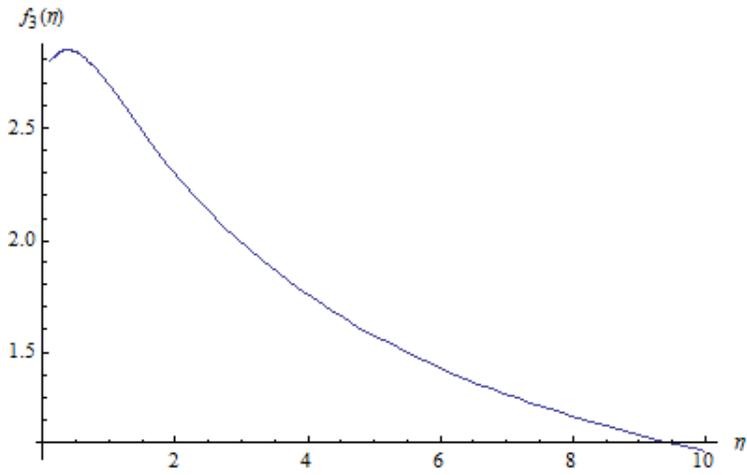}\\
  (c)
  \caption{Plot of the functions: (a)$f_1(\eta)$; (b)$f_2(\eta)$; (c)$f_3(\eta)$.}\label{4}
\end{figure}

In a DW between two opposite parallel states $\mathbf{m(x)}=\pm\hat{z}$,
the magnetization can rotate either in the plane $(\hat{z},\hat{\phi})$,
or in the plane $(\hat{z},\hat{\rho})$. The corresponding two types
of DWs have magnetization $\mathbf{m(x)}=\hat{z}\cos\theta+\hat{\phi}\sin\theta$
or $\mathbf{m(x)}=\hat{z}\cos\theta+\hat{\rho}\sin\theta$ where $\theta$
depends on $z$. These two types of domain walls will be denoted as
P1 and P2, respectively. 
Similarly, in a DW between two opposite vortex states $\mathbf{m(x)}=\pm\hat{\phi}$
the magnetization rotates either in the plane $(\hat{\phi},\hat{z})$,
or in the plane $(\hat{\phi},\hat{\rho})$. The corresponding expressions
for magnetization in these DWs are: $\mathbf{m(x)}=\hat{\phi}\cos\theta+\hat{z}\sin\theta$
and $\mathbf{m(x)}=\hat{\phi}\cos\theta+\hat{\rho}\sin\theta$. We
denote these two types of domain walls as V1 and V2, respectively.
We will take the trial function for $\theta$ to be
\begin{equation}
\theta(z)=2\arctan e^{\frac{z}{l}},\label{trial function}
\end{equation}
where $l$ is the DW width. Plugging the expressions for magnetization
with $\theta(z)$ defined by Eq. \eqref{trial function} into the
energy functionals \eqref{exchange} and \eqref{dip3} and calculating
the integrals, we arrive at the DW energy (the energy of a tube with
a DW minus the energy of the corresponding single domain state) for
the four types of domain walls:
\begin{equation}
E_{P1}=C(\frac{1}{\eta}+\eta+\lambda f_{1}(\eta)),\label{P1}
\end{equation}
\begin{equation}
E_{P2}=C(\frac{1}{\eta}+\eta+\lambda f_{1}(\eta)+\kappa\eta^{2}f_{2}(\eta)),\label{P2}
\end{equation}
\begin{equation}
E_{V1}=C(\frac{1}{\eta}-\eta+\lambda f_{3}(\eta)),\label{V1}
\end{equation}
\begin{equation}
E_{V2}=C(\frac{1}{\eta}+\kappa\eta^{2}f_{2}(\eta)),\label{V2}
\end{equation}
where we have introduced the dimensionless DW width $\eta=\frac{l}{R}$
and dimensionless coordinate $\zeta=\frac{z}{l}$. Other notations
are as follows: $C=4\pi Ad$, $\lambda=\frac{\mu_{0}M_{0}^{2}Rd}{8\pi A}=\frac{Rd}{8\pi\lambda_{x}^{2}}$,
$\kappa=\frac{\mu_{0}M_{0}^{2}R^{3}}{4\pi Ad}=\frac{R^{3}}{4\pi\lambda_{x}^{2}d}$,
\[
f_{1}(\eta)=\int d\zeta d\zeta'kK(k)\mathrm{sech}^{2}\zeta\mathrm{sech}^{2}\zeta',\;
\]
\[
f_{2}(\eta)=\int d\zeta d\zeta'[kK(k)-k_{d}K(k_{d})]\mathrm{sech}\zeta\mathrm{sech}\zeta',
\]
\[
f_{3}(\eta)=\int d\zeta d\zeta'kK(k)\tanh\zeta\mathrm{sech}\zeta\tanh\zeta'\mathrm{sech}\zeta',
\]
where $k=\left[1+\frac{1}{4}\eta^{2}(\zeta-\zeta')^{2}\right]^{-1/2}$,
$k_{d}=\left[1+\frac{d^{2}}{4R^{2}}+\frac{1}{4}\eta^{2}(\zeta-\zeta')^{2}\right]^{-1/2}$,
and $K(k)$ is the complete elliptic integral of the first kind. The
function $f_{1}(\eta)$ decreases monotonically as $\eta$ increases,
$f_{3}(\eta)$ has a maximum at $\eta=0.37$, and $f_{2}(\eta)$ has
a maximum that depends on the value of $\frac{d}{R}$ (for $\frac{d}{R}=0.1$
it is at $\eta=2.9$). All functions $f_{i}(\eta)$, $i=1,2,3$ are
positive for any $\eta$ and turn into zero at $\eta\rightarrow\infty$.
They are plotted in Fig. \ref{4}.

From the positivity of $f_{i}(\eta)$, it follows that $E_{P1}$,
$E_{P2}$ and $E_{V2}$ are positive since each term in Eqs. \eqref{P1},
\eqref{P2} and \eqref{V2} is positive. For the V1 case, there is
one negative term $-C\eta$. Any solution (if it does exist) of the equation for extremum of the
energy \eqref{V1}, i.e. $\lambda f_{3}^{\prime}(\eta)=\eta^{-2}+1$, can only lie between 0 and 0.37 since only in this region is $f_{3}^{\prime}(\eta)$ positive. In the same
interval the difference $1/\eta-\eta$ is positive and therefore the
energy $E_{V2}$ is also positive. Thus, if the DW energy has a minimum,
it must be at a value of $\eta$ smaller than 0.37. Since it was already
proved that $E_{V1}>0$ for $\eta<1$, we conclude that inner domain
walls are never stable. However, they can exist as metastable states.
For the P1 DW and $\lambda=0$ (negligible dipolar interaction), the
minimum is realized at $\eta=1$, i.e. at $l=R$. So, if only the
exchange energy is considered, the DW width is equal to the radius
of the tube. If the dipolar energy is included, the minimum shifts
to $l>R$. This looks reasonable, since the volume charge $\rho_{M}(\mathbf{x})=\sin\theta\partial_{z}\theta$
inside the DW has the same sign throughout the wall and it tends to
spread out the DW width to reduce the dipolar energy. The energy $E_{P2}$
differs from $E_{P1}$ only by a positive term $\kappa\eta^{2}f_{2}(\eta)$.
Thus, the P2 DW is always less favorable than the P1 one. For a V1
DW, numerical calculations show that $E_{V1}$ has a minimum only
at $\lambda>97.109$. The value of $\eta$ realizing the minimum of
energy at $\lambda=97.109$ is $0.2467$. It monotonically decreases
with increasing $\lambda$. Thus, the metastable V1 DW exists at $\lambda>97.109$
and its width is smaller than $0.2467R$.

Summarizing, under our assumptions no stable DW exist in the middle
of the magnetic tube, but they can exist as metastable configurations.
It is well known that dipolar interaction generates a stripe domain
structure in a bulk rectangular ferromagnetic slab. It does not happen
in a thin-walled long magnetic tube. The domains in it can not appear
as equilibrium state, but only as a result of the growth process
from two or more nuclei. However, the edge DW regularly appear. (see
section \ref{half DW})

\section{MAGNETIC STRUCTURES IN EXTERNAL MAGNETIC FIELD}

The external magnetic field $\mathbf{H}$ interacts with magnetization
via the Zeeman energy:
\begin{equation}
E_{Z}=-\mu_{0}M_{0}\int d^{3}x\mathbf{m(x)}\cdot\mathbf{H}.\label{Zeeman}
\end{equation}
If the external field is directed along $z$-axis (parallel field),
the azimuthal symmetry is still retained. A field of any other direction
destroys this symmetry. We will first consider the parallel field
and then the field along 
a transverse direction.

\subsection{Parallel field}

\begin{figure}
  \centering
  \includegraphics[width=0.6\textwidth]{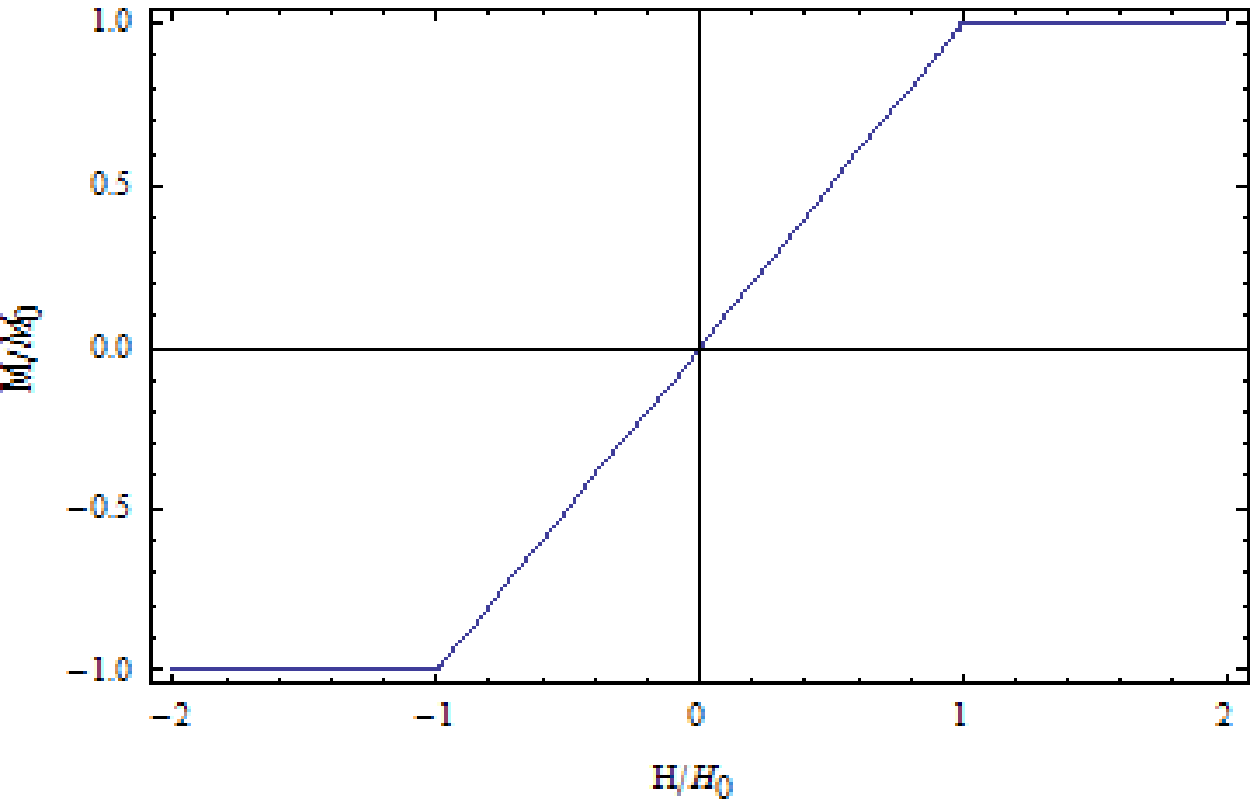}\\
  (a)\\
  \includegraphics[width=0.6\textwidth]{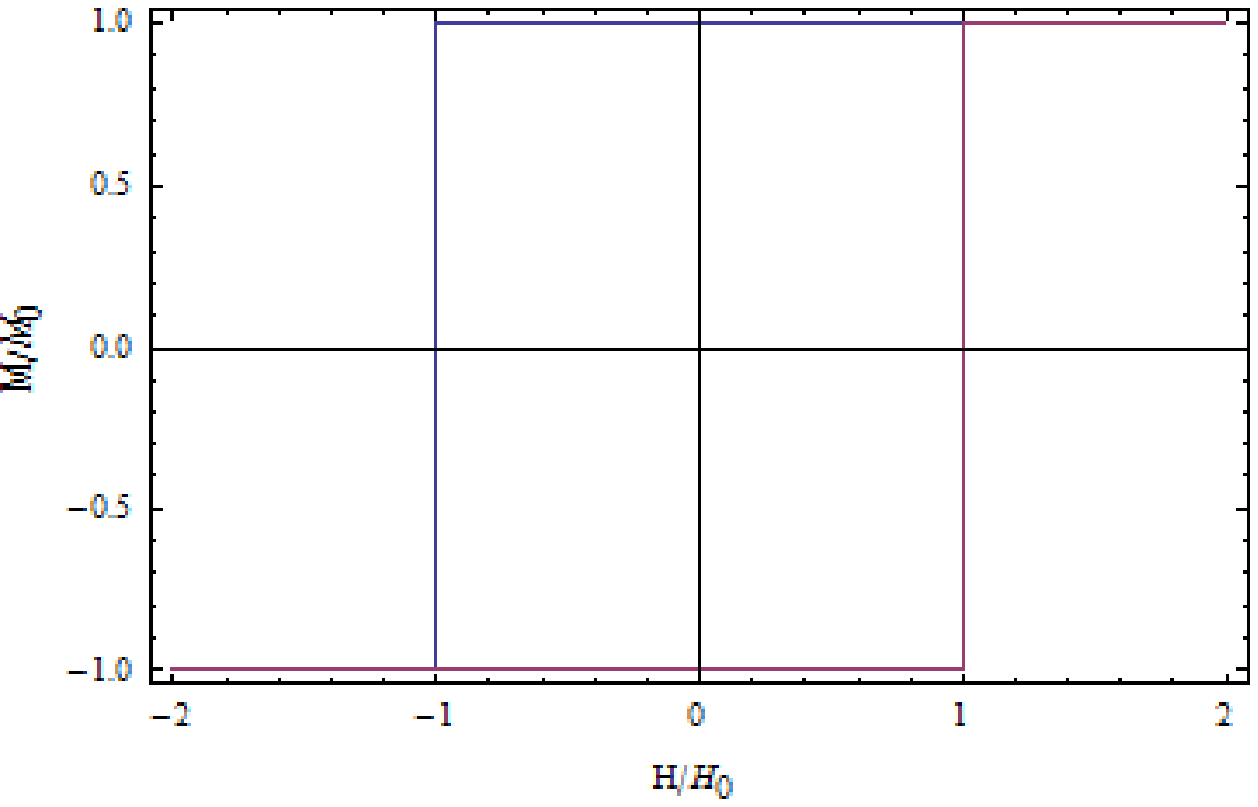}\\
  (b)
  \caption{(a) The dependence of magnetization in a parallel magnetic field at $E_P>E_V$. Note the absence of hysteresis. (b) Hysteresis in a parallel external field at $E_P<E_V$.}\label{7}
\end{figure}

First we analyze the action of the parallel magnetic field $\mathbf{H}=H\hat{z}$
onto the vortex state. The magnetization acquires a finite $z$-component
and can be written as
\begin{equation}
\mathbf{m(x)}=\hat{\phi}\cos\theta+\hat{z}\sin\theta,\label{mp}
\end{equation}
where $\theta$ is a function of $H$ only. At magnetic field $H$
varying from zero to some critical value $H_{c}$, the angle $\theta$
changes from zero to $\pm\pi/2$ and the configuration changes continuously
from the vortex to parallel state. Eq. \eqref{Zeeman} implies that
$E_{Z}=-\mu_{0}M_{0}\int d^{3}xH\sin\theta=-2\pi\mu_{0}M_{0}RdLH\sin\theta\triangleq-E_{Z0}\sin\theta$,
where $E_{Z0}=2\pi\mu_{0}M_{0}RdLH$ is the absolute value of $E_{Z}$
for the parallel states. The exchange energy of the state \eqref{mp}
is equal to the energy of the vortex state at zero field multiplied
by the factor $\cos^{2}\theta$. The dipolar energy of the configuration
\eqref{mp} is equal to the dipolar energy of the parallel state at
$H=0$ multiplied by $\sin^{2}\theta$. The total energy of configuration
\eqref{mp} is
\begin{eqnarray}
E(\theta)&=&E_{V}\cos^{2}\theta+E_{P}\sin^{2}\theta-E_{Z0}\sin\theta\nonumber\\
&=&(E_{P}-E_{V})\left[\sin\theta-\frac{E_{Z0}}{2(E_{P}-E_{V})}\right]^{2}+\frac{E_{Z0}^{2}}{4(E_{P}-E_{V})}+E_{V}.\label{}
\end{eqnarray}
If $E_{P}>E_{V}$ (the tube is in the vortex state at $h=0$) and
$E_{Z0}<2\left(E_{P}-E_{V}\right)$, then $\sin\theta=\frac{E_{Z0}}{2(E_{P}-E_{V})}$
corresponds to the minimum of energy, and the equilibrium energy is
between the energies of V and P states. The plot of the net magnetization
$M=M_{0}\sin\theta$ vs. $H$ is shown in Fig. \ref{7}(a). The critical field
at which the parallel state is reached is $H_{C}=\frac{E_{P}-E_{V}}{\pi\mu_{0}M_{0}RdL}$.
If $E_{P}<E_{V}$ (the tube is in the parallel state at $H=0$), then
$\sin\theta=\frac{E_{Z0}}{2(E_{P}-E_{V})}$ corresponds to the maximum
of energy that realizes an energy barrier between the two parallel
states. The plot $M$ vs. $H$ displays a rectangular hysteresis loop
with the coercive force equal to $H_{c}$, as shown in Fig. \ref{7}(b).

This feature reminds the Stoner-Wohlfarth model \cite{Stoner-Wohlfarth}.
The reason of the hysteresis in the Stoner-Wolfarth model is the crystal
field anisotropy, i.e. spin-orbit interaction. In magnetic tubes the
geometry is anisotropic. At $E_{P}>E_{V}$ and magnetic field along
the axis of the tube, the geometric anisotropy keeps the component
of magnetization perpendicular to the field and provides a continuous
transition at a critical field to completely parallel magnetization.
In the opposite case $E_{P},E_{V}$ and initial magnetization antiparallel
to the field, an intrinsic hysteresis appears in the absence of the
spin orbit interaction due to the exchange and dipolar forces. At
magnetic field $\pm H_{c}$ the magnetization flips. 

We considered the mechanism of a coherent magnetization flip resulting
in a comparatively large coercive force. 
Other mechanisms were proposed in \cite{equilibrium states,reversal modes}.
The only mechanism relevant to the limit of a long thin tube is the
propagation of a vortex domain wall between two opposite parallel
configuration. The vortex DW propagation is favorable if the dipolar
energy dominates. in Ref. \cite{equilibrium states}, Vortex DW propagation
is also described and argued to be closely related to the half vortex
DWs at the tube ends for one parallel state. Notice that in \cite{equilibrium states}
they also have an almost rectangular-shaped hysteresis loop. While
transverse DW propagation is favorable if the exchange energy dominates
such that a transverse DW has lower energy than a vortex DW \cite{reversal modes}. We do not take pinning into account. It is expected that if pinning is included the hysteresis loop will have a larger coercive force.


\subsection{Infinite magnetic tube in a transverse magnetic field}

\label{transverse external field} Now we consider the transverse
magnetic field applied to a tube that was initially in a vortex state.
Let the field be in the $x$ direction: $\mathbf{H}=H\hat{x}$. The
field violates the azimuthal symmetry. The direction of magnetization
in the transverse field deviates from $\hat{\phi}$. We assume that
$\mathbf{m(x)}$ depends only on $\phi$ and it does not have $z$-component:
\begin{equation}
\mathbf{m(x)}=\hat{\phi}\cos\theta(\phi)+\hat{\rho}\sin\theta(\phi).\label{MT}
\end{equation}
At $H=0$ the vortex state $\mathbf{m(x)}=\hat{\phi}$ has minimal
energy. At very large $H$ the transverse state $\mathbf{m(x)}=\hat{x}$
is energy favorable. These two states are topologically different:
the former has the winding number 1 while the latter has the winding
number 0. The transition from one of them to another can not proceed
continuously. It should be a first order transition at a critical
value of magnetic field accompanying with the discontinuity of net
magnetization $M$. It was reported in \cite{bland review} that such
a transition indeed happens in a ferromagnetic ring: at a critical
transverse field, the magnetization transits from the vortex state
to the onion state. We will show that the magnetization in a tube
also follows this scenario. The exact transition behavior to the onion
state is still unknown. It may include the escape of the magnetization
from $xy$-plane. Below we compare the energy of the modified vortex
state and that of the onion state to determine the critical field
of the transition. 

The energy of the onion state is the sum of the onion energy $E_{O}$
at zero magnetic field and the Zeeman energy $E_{Z}=-\mu_{0}M_{0}\int d^{3}xH\cos\theta=-\pi\mu_{0}M_{0}HRdl[1+J_{0}(2a)]$.
(We again use the trial function $\theta(\phi)=-a\sin(2\phi)$.) The
total onion energy is:
\begin{eqnarray}
E_{O}(a,H)&=&4\pi A\frac{Ld}{R}a^{2}+\frac{\pi}{2}\mu_{0}M_{0}^{2}RdL[1-J_{1}(2a)]-\pi\mu_{0}M_{0}HRdl[1+J_{0}(2a)]\nonumber\\
&=&4\pi A\frac{Ld}{R}\left\{ a^{2}+\frac{R^{2}}{8\lambda_{x}^{2}}\left[1-J_{1}(2a)-2\frac{H}{M_{0}}\left[1+J_{0}(2a)\right]\right]\right\}. \label{}
\end{eqnarray}
This energy should be minimized with respect to $a$. Let the equilibrium
value of $a$ be $a_{0}$. It is physically obvious that the equilibrium
value of variational parameter $a_{0}$ monotonically decreases with
$H$ increasing and reaches zero at $H=\infty$. Therefore, it cannot
be larger than $0.92$ -- the maximum value of $a_{0}$ at zero field.

The vortex state also changes in the presence of magnetic field. The
local magnetization approaches to the direction of magnetic field,
so that the vortex configuration acquires a non-zero total magnetic
moment. We describe this changes by a trial function:
\begin{equation}
\mathbf{m(x)}=\frac{\hat{\phi}+b\hat{x}}{\sqrt{1+b^{2}-2b\sin{\phi}}}=\hat{\phi}\frac{1-b\sin\phi}{\sqrt{1+b^{2}-2b\sin{\phi}}}+\hat{\rho}\frac{b\cos\phi}{\sqrt{1+b^{2}-2b\sin{\phi}}},\label{}
\end{equation}
where $b$ is a non-negative parameter which does not depend on $\phi$.
At $b=0$ the trial configuration turns into the genuine vortex state
and at $b=\infty$ it turns into the transverse state. The energy
of the modified vortex state reads:
\begin{eqnarray}
E_{MV}(b,H)&=&\pi A\frac{Ld}{R}\frac{2-b^{2}}{1-b^{2}}+\frac{\pi}{4}\mu_{0}M_{0}^{2}RdLb^{2}
-\frac{2\mu_{0}M_{0}HRdl}{b}[(1+b)E(k)-(1-b)K(k)]\nonumber\\
&=&\pi A\frac{Ld}{R}\left\{ \frac{2-b^{2}}{1-b^{2}}+\frac{R^{2}}{\lambda_{x}^{2}}\left[\frac{b^{2}}{4}-\frac{2}{\pi}\frac{H}{M_{0}}\left[\frac{b+1}{b}E(k)-\frac{1-b}{b}K(k)\right]\right]\right\} \label{MV},
\end{eqnarray}
where $k=\frac{\sqrt{4b}}{1+b}$; $K(k)$ and $E(k)$ are complete
elliptic integrals of the first and the second kind, respectively.
The energy \eqref{MV} should be minimized with respect to $b$. Let
the equilibrium value of $b$ be $b_{0}$.

\begin{figure}
  \centering
  \includegraphics[width=0.6\textwidth]{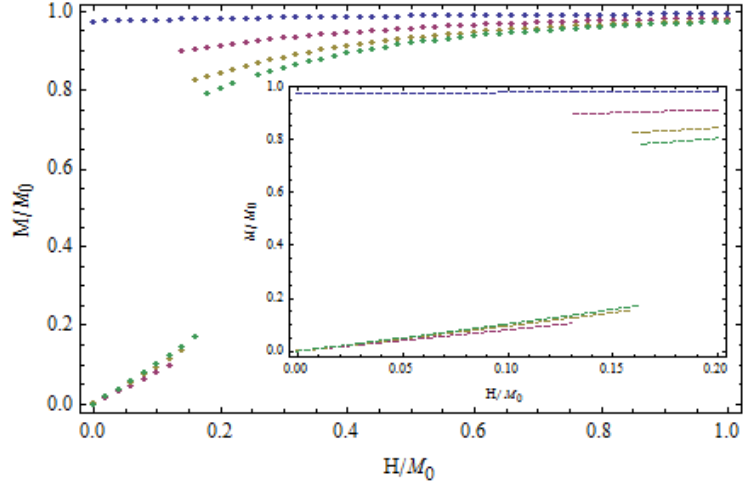}\\
  \caption{(Color Online) Plot of magnetization vs. magnetic field (in units $M_0$) in a transverse field for 4 values of the parameter $R/\lambda =\infty, 8, 4, 2$ (from bottom to top). The inset shows $M$ vs. $H$ in small $H$ region. $H_C/M_0$ for $R/\lambda =\infty, 8, 4, 2$ are 0.163, 0.159 and 0.131, respectively.}\label{9}
\end{figure}

\begin{figure}
  \centering
  \includegraphics[width=0.6\textwidth]{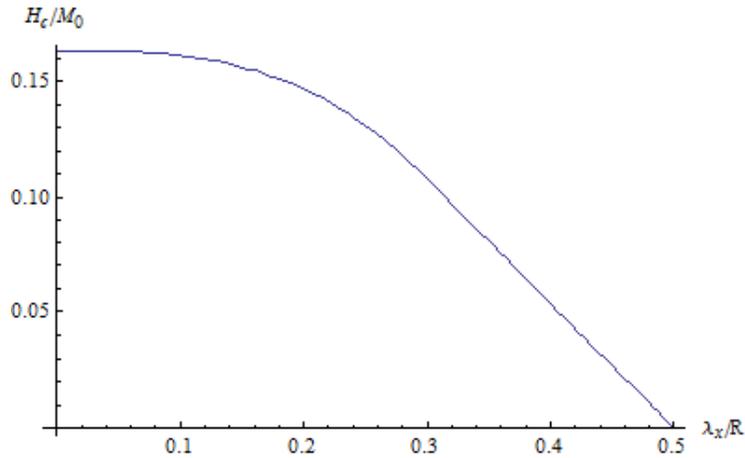}\\
  \caption{Plot of the critical field $H_C$ vs. $\lambda_x/R$.}\label{10}
\end{figure}

Given $a_{0}$ and $b_{0}$, the phase transition condition $E_{MV}(b_{0})=E_{O}(a_{0})$
determines the critical field $H_{C}$ (the field at which transition between the onion and the modified vortex state takes place). The basic equations of the
Table 1 show that, at $\frac{R}{\lambda_{x}}<2$, the transverse state
has lower energy than the vortex state at zero field, and the more
at any non-zero field. Thus, the transition proceeds only at $\frac{R}{\lambda_{x}}>2$.
Fig. \ref{9} presents the plots of $M$ vs. $H$ (both divided by $M_{0}$)
for 4 different values of $\frac{R}{\lambda_{x}}=\infty,8,4,2$. The
corresponding values of the critical field $\frac{H_{C}}{M_{0}}$
are $0.163$, $0.159$ and $0.131$. At $\frac{R}{\lambda_{x}}=2$
the tube is in the onion state at zero field. Fig. \ref{10} presents the
critical field $H_{c}$ v.s. $\lambda_{x}$. Note that these results does not depend on the ratio $d/R$, although we still have to be in the $\frac{d}{R}\ll 1$ limit.

\section{THE EDGE MAGNETIC TEXTURES}\label{half DW}

Although we have proved that the DW in the middle
of a tube is never energy favorable, it occurs that a finite tube
in the parallel state has a vortex-like magnetization at its edges.
The transitional texture from the parallel to vortex magnetization
can be imagined as half of the P1 type DW. We will call such a texture
the edge DW (EDW). Such a structure was found theoretically in Ref.
\cite{equilibrium states} where it is called a mixed state. The EDW
weakens the stray magnetic field near the edges and reduces the dipolar
energy. From the magnetic charge point of view, in the EDW the surface
charge at the edges is spread out into the volume, so that the total
charge remains invariant. When the reduction of dipolar energy exceeds
the increase of the exchange energy, the state with the EDW becomes
stable. Since the total length $L$ is much larger than the width
$l$ of the EDW, the magnetization in the tube beyond the two EDW
does not change.

Let us consider the EDW in more details. We consider an semi-infinite
tube whose axis coincides with the positive half-axis $z$. We also
assume that the stable magnetic configuration is parallel at $z\rightarrow\infty$.
The magnetization at arbitrary $z$ can be represented as $\mathbf{m(x)}=\hat{z}\cos\theta+\hat{\phi}\sin\theta$,
where $\theta$ is a function of $z$ satisfying the boundary conditions
$\theta(\infty)=\pi$, while $\theta(0)$ is a variational parameter.
If either it occurs not equal to $\pi/2$ or $\frac{d\theta}{dz}\neq0$,
it means that the surface magnetic charge does not vanish completely.
To take in account this effect, we employ the trial function:
\begin{equation}
\theta(z)=2\arctan e^{\frac{z+z_{0}}{l}},
\end{equation}
where $z_{0}=\log\tan\frac{\theta_{0}}{2}$.%
\footnote{In Ref. \cite{equilibrium states} a linear trial function was used.%
} This trial function is reasonable also in the presence of external
magnetic field parallel to the axis. Therefore, we will perform variational
calculations for arbitrary magnetic field $H$. For this purpose we
need to complement the Hamiltonian by the Zeeman energy: 
$E_{Z}=-\mu_{0}M_{0}\int d^{3}x\mathbf{H}\cdot(\mathbf{m(x)}+\hat{z})=\mu_{0}M_{0}H\int d^{3}x(1+\cos\theta)$
where $\mathbf{H}=-H\hat{z}$. Calculations similar to the case of
the P1 type DW result in the following energy $E_{E}$ of the edge
configuration:
\begin{eqnarray}
E_{E}&=&\frac{1}{2}C\{ (1-\tanh\zeta_{0})(\frac{1}{\eta}+\eta)\nonumber\\
&&+\lambda_{E}\left[2\tanh\zeta_{0}g(\eta,\zeta_{0})+f(\eta,\zeta_{0})+(\tanh^{2}\zeta_{0}-1)\log(\frac{8R}{d}+\frac{3}{2})\right]\nonumber\\
&&+\mu\eta\left[\log(2\cosh\zeta_{0})-\zeta_{0}\right]\} ,\label{}
\end{eqnarray}
where $\lambda_{E}=2\lambda$, $\mu=\frac{\mu_{0}M_{0}HR^{2}}{A}=\frac{R^{2}}{\lambda_{x}^{2}}\frac{H}{M_{0}}$, 
$\zeta_{0}=\frac{z_{0}}{l}$,
\[
f(\eta,\zeta_{0})=\int_{0}^{\infty}\int_{0}^{\infty}d\zeta d\zeta'kK(k)\mathrm{sech}^{2}(\zeta+\zeta_{0})\mathrm{sech}^{2}(\zeta'+\zeta_{0}),
\]
\[
g(\eta,\zeta_{0})=\int_{0}^{\infty}d\zeta kK(k)|_{\zeta'=0}\mathrm{sech}^{2}(\zeta+\zeta_{0})\mathrm{sech}^{2}(\zeta'+\zeta_{0}).
\]
Other notations are the same as in section \ref{DW}.

\begin{figure}
  \centering
  \includegraphics[width=0.6\textwidth]{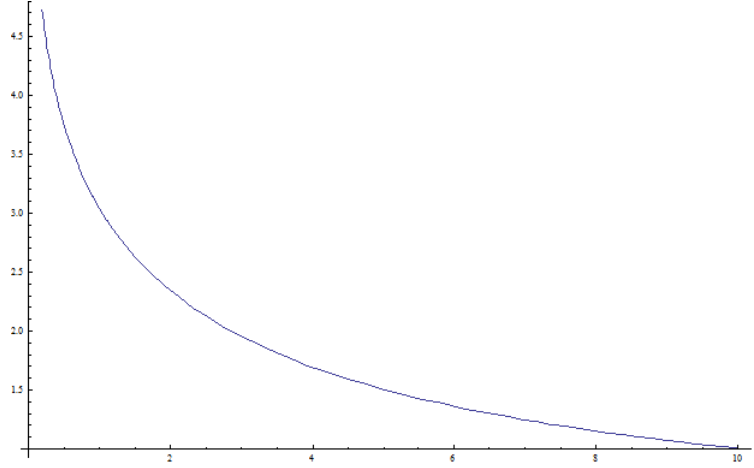}\\
  \caption{Plot of the function $f(\eta,0)$.}\label{11}
\end{figure}

\begin{figure}
  \centering
  \includegraphics[width=0.6\textwidth]{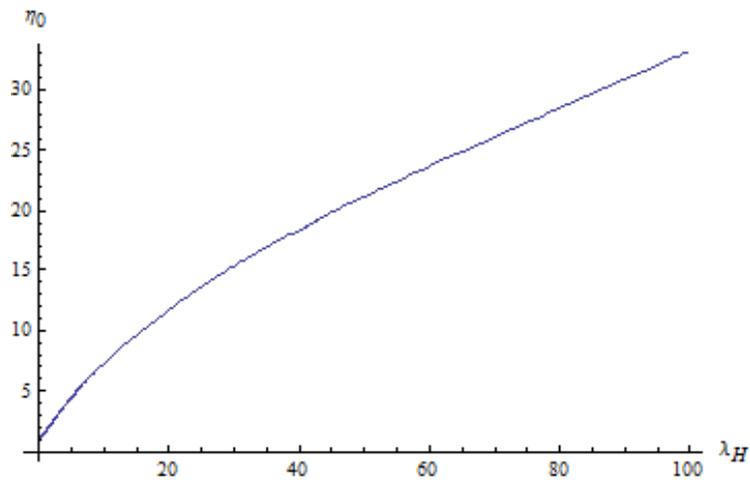}\\
  \caption{Plot of $\eta_0$ vs. $\lambda_E$ for $d/R=0.1$.}\label{12}
\end{figure}

\begin{figure}
  \centering
  \includegraphics[width=0.6\textwidth]{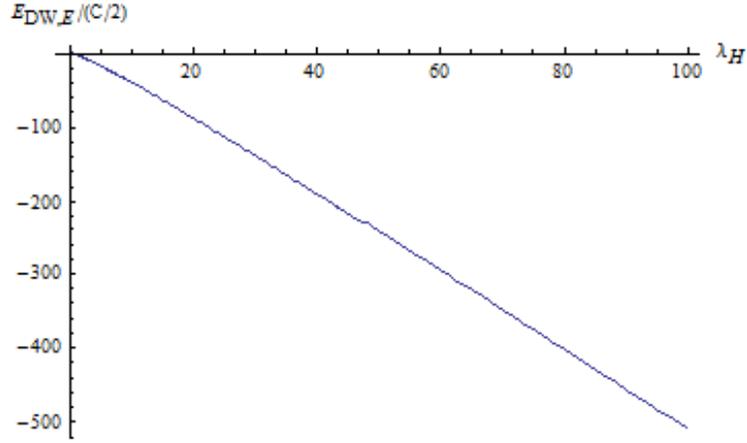}\\
  \caption{Plot of $E_{DW,E}$ vs. $\lambda_E$ at $d/R=0.1$.}\label{13}
\end{figure}

\begin{figure}
  \centering
  \includegraphics[width=0.6\textwidth]{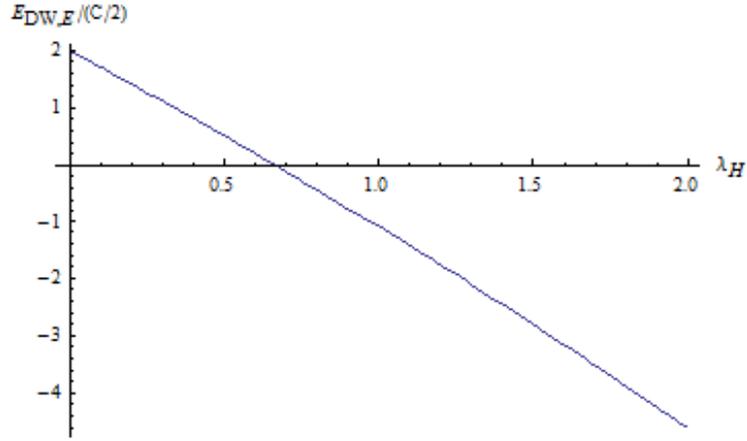}\\
  \caption{Plot of $E_{DW,E}$ vs. $\lambda_E$ at $d/R=0.1$, small $\lambda_E$ region. $E_{DW,E}=0$ at about $\lambda_E=0.66$.}\label{14}
\end{figure}

Let us start with the analysis of the EDW at zero magnetic field.
The EDW is stable if $E_{E}<0$. In turn this condition implies a
necessary condition of the EDW stability $f(\eta,0)<\log\frac{8R}{d}+\frac{3}{2}$.
In Fig. \ref{11} we plot the graph of $f(\eta,0)$ vs. $\eta$ for $\eta\in[0.1,10]$.
The necessary condition of the EDW stability is satisfied for $\eta>1$.
Since $f(\eta,0)$ monotonically decreases, the value $\eta=\eta_{0}$
which minimizes $E_{E}$ is also located in the region $\eta>1$.
Thus, the stable EDW has the width $l>R$. Fig. \ref{12} presents the graph
of $\eta_{0}$ vs. $\lambda_{E}$. It shows that, at $\lambda_{E}$
increasing from zero, $\eta_{0}$ increases from 1 monotonically.
Asymptotically at large $\lambda_{E}$, the dependence of $\eta$
on $\lambda_{E}$ becomes linear. 
Figs. \ref{13} and \ref{14} present the graphs of $E_{E}/(C/2)$ vs. $\lambda_{E}$
for large and small $\lambda_{E}$, respectively. The calculations
were performed at a fixed value $\frac{d}{R}=0.1$. Fig. \ref{14} shows
that the energy becomes negative at $\lambda_{E}>0.66$. Since $\lambda_{E}=2\lambda=\frac{Rd}{4\pi\lambda_{x}^{2}}$,
the criteria of the EDW stability at $\frac{d}{R}=0.1$ is $Rd>2.64\pi\lambda_{x}^{2}$,
i.e. $R>9.1\lambda_{x}$. For typical figures $R\sim100nm$, $d\sim10nm$
and $\lambda_{x}$ several nanometers, the stability condition is
readily satisfied.

The value $\zeta_{0}$ at zero magnetic field is not zero, but it
is rather small (about 0.04). However it grows with increasing magnetic
field as it is shown in the Table 2. The same table displays the equilibrium
values parameter $\eta$ and energy of the edge texture as function
of magnetic field ($\mu$) at $d/R=0.1$ and $\lambda_{E}=50$. 
The Table 2 also shows the dependence of the equilibrium value of
$\eta$ on $\mu$. Magnetic field reduces $\eta$ to values less than
one, i.e. it narrows the edge domain wall width to $l<R$. The last
column of the Table 2 shows that the energy of the EDW becomes positive
at $\mu>427$. The field $H$ corresponding to $\mu=427$ is $0.0680M_{0}$.
Thus, in a tube with $\frac{d}{R}=0.1$ and $\lambda_{E}=50$, the
EDW vanishes at a critical field $H_{E}=0.0680M_{0}$.

Note that the critical field $H_E$ here is obtained by equalizing the energy of the parallel and the EDW state. If we want to examine the nucleation of EDW in more detail, we should perform stability analyses for both the parallel and the EDW state, and we are likely to get a hysteresis between the two states. Nevertheless the field $H_{E}$ here remains a good estimate of the transition field between these two states.

\bigskip{}
Table 2. Energies and parameters of EDW in external field at $d/R=0.1$
and $\lambda_{E}=50$.

\smallskip{}
\makebox[1\textwidth]{%
\begin{tabular}{|c|c|c|c|}
\hline
$\mu$  & $\eta$  & $\zeta_{0}$  & $E_{DW,E}/(C/2)$\\
\hline
0  & 21.7496  & 0.0385034  & -242.641\\
\hline
100  & 0.792432  & 0.108815  & -75.4210\\
\hline
200  & 0.408263  & 0.139860  & -39.8710\\
\hline
300  & 0.278412  & 0.168046  & -18.9154\\
\hline
400  & 0.229659  & 0.230356  & -3.68804\\
\hline
500  & 0.213332  & 0.285309  & 9.35049\\
\hline
600  & 0.206421  & 0.320128  & 21.4643\\
\hline
\end{tabular}%
} \bigskip{}

\section{STRAY MAGNETIC FIELD}

\begin{figure}
  \centering
  \includegraphics[width=0.6\textwidth]{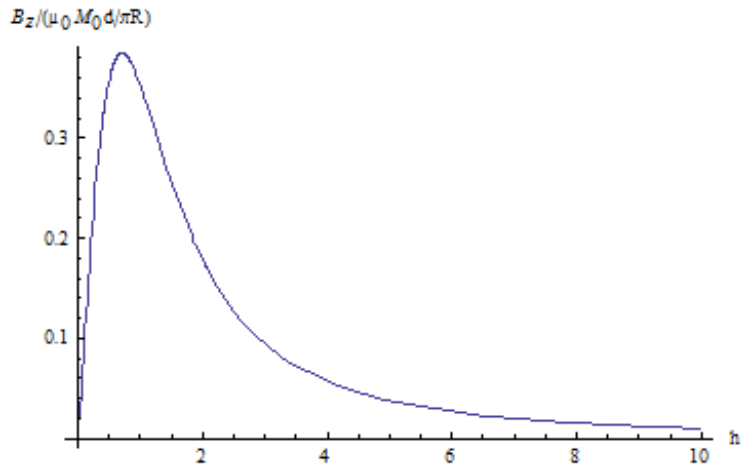}\\
  \caption{Plot of $B_z(\mathbf{x})$ (divided by $\frac{\mu_0M_0d}{2R}$) vs. $z$, which presents the field along the $z-$axis.}\label{15}
\end{figure}

\begin{figure}
  \centering
  \includegraphics[width=0.6\textwidth]{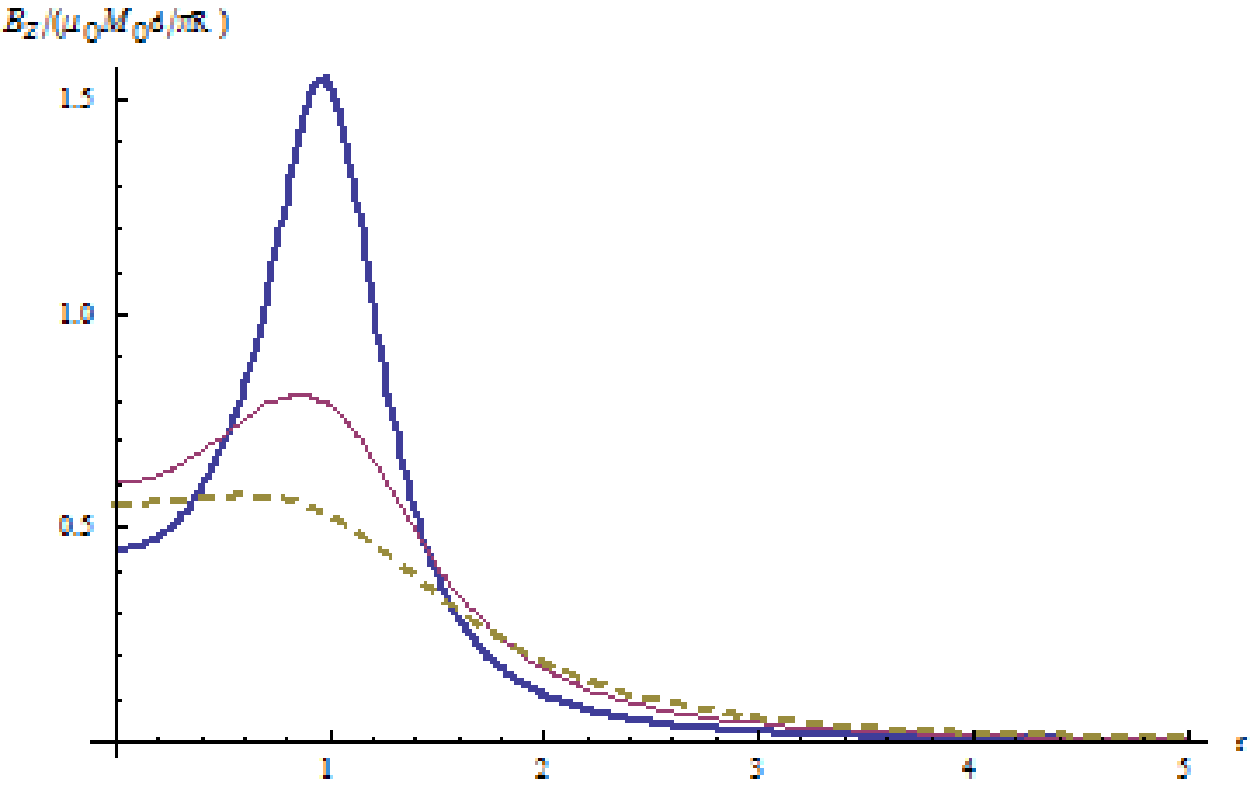}\\
  (a)
  \includegraphics[width=0.6\textwidth]{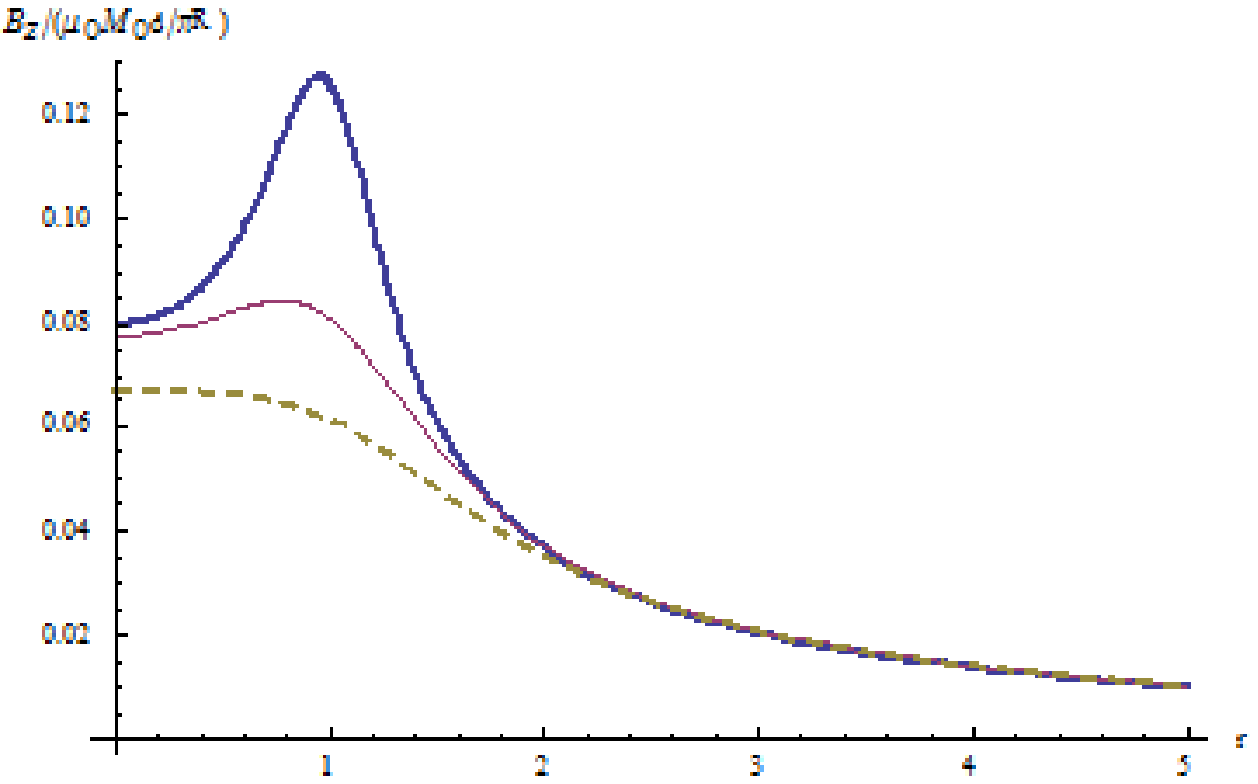}\\
  (b)
  \caption{(Color online) (a) Plot of $B_z(\mathbf{x})$ (divided by $\frac{\mu_0M_0d}{\pi R}$) vs. $r$ at $h=1/3$ (thick), 2/3 (solid), 1(dashed), respectively, for a tube in the parallel state. (b) The same as (a), but for a tube with EDW. The parameters describing the EDW are $\eta=21.7496$ and $\zeta=0.0385034$.}\label{16}
\end{figure}

Now we calculate the field produced by the nanotube. The field at
a point $\mathbf{x}$ outside the tube reads:
\begin{equation}\label{fieldproduced}
\mathbf{B(x)}=-\frac{\mu_{0}}{4\pi}M_{0}\nabla\phi(\mathbf{x})=-\frac{\mu_{0}}{4\pi}M_{0}\mathbf{\nabla_{x}}[\int d^{3}x'\frac{\rho_{M}(\mathbf{x'})}{r}+\int dA'\frac{\sigma_{M}(\mathbf{x'})}{r}].
\end{equation}
For a vortex state, $\rho_{M}(\mathbf{x})=\sigma_{M}(\mathbf{x})=0$
everywhere, so that $\mathbf{B(x)}=0$ everywhere outside the material.
In the parallel state $\mathbf{m(x)}=\hat{z}$, the only charges are
$\sigma_{M}(\mathbf{x})=1$ at $z=\frac{L}{2}$ and $\sigma_{M}(\mathbf{x})=-1$
at $z=-\frac{L}{2}$. If the observation point is not too close to
the tube, it is possible to neglect the change of radius $\rho^{\prime}$
replacing it by $R$. Then integration over the radius is reduced
to multiplication by $d$. Integration over $\phi$ leads to complete
elliptic integrals. We are mostly interested in the $z$ component
of the field, which is:
\begin{equation}
B_{z}(\mathbf{x})=\frac{\mu_{0}}{\pi}M_{0}Rd\left[\frac{(z-z')E(\sqrt{\frac{4\rho R}{(\rho+R)^{2}+(z-z')^{2}}})}{\sqrt{(\rho+R)^{2}+(z-z')^{2}}[(\rho-R)^{2}+(z-z')^{2}]}\right]_{z=-\frac{L}{2}}^{z=\frac{L}{2}}.\label{48}
\end{equation}
For an observation point separated from the tube by a distance much
larger than $R$, the field is well approximated by the field of two
point magnetic charges. In regions near the edges magnetic field changes
rapidly. Let us consider for example a region $z-\frac{L}{2}\ll L$.
In a part of this region the finiteness of the radius $R$ is important.
The contribution to the field from the edge $z=-\frac{L}{2}$ can
be ignored. Introducing dimensionless variables $h=\frac{z-\frac{L}{2}}{R}$
and $r=\frac{\rho}{R}$, we find
\begin{equation}
B_{z}(\mathbf{x})=\frac{\mu_{0}M_{0}d}{\pi R}\frac{hE(\sqrt{\frac{4r}{(r+1)^{2}+h^{2}}})}{\sqrt{(r+1)^{2}+h^{2}}[(r-1)^{2}+h^{2}]}.\label{49}
\end{equation}
Eq. (23) becomes simpler on the $z-$axis: $B_{z}(0,z)=\frac{\mu_{0}M_{0}d}{2R}\frac{h}{(1+h^{2})^{\frac{3}{2}}}$.
By symmetry the field on the $z-$axis must be parallel to the $z-$axis.
Fig. \ref{15} presents the graph of $B_{z}(0,z)$ divided by $\frac{\mu_{0}M_{0}d}{2R}$
vs. $h$. It starts from zero, reaches a maximum equal to $\frac{2\sqrt{3}}{9}$
at $h=\frac{\sqrt{2}}{2}$. and then decreases to zero. Thus, the
maximum field along the $z$ axis locates at $\frac{2\sqrt{3}}{9}R\approx0.385R$
above the upper edge. Fig. \ref{16}(a) presents the graphs of $B_{z}(\mathbf{x})$
in units $\frac{\mu_{0}M_{0}d}{\pi R}$ vs. $r$ at different $h$.
Note the maxima of $B_{z}(\mathbf{x})$ at $r\approx1$, i.e. at $\rho\approx R$.
The smaller is $h$, the larger and sharper is the maximum. It is
well-known singularity of the field of homogeneously charged ring.

What about field produced by an EDW state? The EDW near the $z=\frac{L}{2}$ edge is described like before: $\mathbf{m(x)}=\hat{z}\cos\theta+\hat{\phi}\sin\theta$, where \begin{equation}
\theta(z)=2\arctan e^{\frac{z-\frac{L}{2}-z_{0}}{l}}.
\end{equation}
Using Eq. \eqref{fieldproduced}, the $z-$component of magnetic field at a point $(\rho,\phi,z)$ above the $z=\frac{L}{2}$ edge is:
\begin{eqnarray}\label{}
B_z(r,h)&=&\frac{\mu_0M_0d}{\pi R}[\frac{hE(\sqrt{\frac{4r}{(r+1)^{2}+h^{2}}})}{\sqrt{(r+1)^{2}+h^{2}}[(r-1)^{2}+h^{2}]}\tanh\zeta_0\nonumber\\
&&+\int_{-\infty}^{0}d\zeta'\frac{hE(\sqrt{\frac{4r}{(r+1)^{2}+(h-\eta\zeta')^{2}}})}{\sqrt{(r+1)^{2}+(h-\eta\zeta')^{2}}[(r-1)^{2}+(h-\eta\zeta')^{2}]}\mathrm{sech}^2(\zeta'-\zeta_0)],
\end{eqnarray}
where in addition to $h$ and $r$ we have defined $\eta=l/R$, $\zeta=\frac{z-\frac{L}{2}}{l}$ and $\zeta_0=z_0/l$. 
To calculate numerically, we use $\eta=21.7496$ and $\zeta_0=0.0385034$ which are values at $d/R=0.1, \lambda_E=50$ and zero external field. Fig. \ref{16}(b) presents $B_z(r,h)$ in units $\frac{\mu_0M_0d}{\pi R}$ at $h=1/3,2/3$ and 1. Comparing with Fig. \ref{16}(a), we see that although peaks remain at $r\approx1$ they are are more than 10 times lower, and also are more smeared. Thus the EDW reduces the field and smears the singularity (though does not completely remove it).

Now we consider the field produced by a periodic array of tubes. Such
an array appears in the experiments employing the template technique,
e.g in \cite{Nielsch 3} and \cite{electrodeposition}. Let the infinite
array of tubes spread out in the $x-y$ plane forming equilateral
triangular lattice (like in the experiment \cite{Nielsch 3}). We
assume that all the tubes are in the $\mathbf{m(x)}=\hat{z}$ parallel
state. Below we calculate $B_{z}(\mathbf{x})$ near the upper plane
of the system $z=\frac{L}{2}$ for the lattice constant $D=3R$ and $2R$. The
field is periodic with periodicity of the lattice. It allows us to
restrict calculations by one quarter of the elementary cell.

Let us introduce a frame of reference with the $z-$axis parallel
to the tube axes and the $x-$axis along one of directions connecting
axes of nearest neighboring tubes, i.e. along one of basic vectors
of the Bravais lattice. The expression for $B_{z}(\mathbf{x})$ is
an infinite sum of the fields produced by a single tube at a particular
place, each of them given by Eq. \eqref{49} with the position of
a tube taking into account. To perform the summation approximately,
we calculate exactly the contribution of several nearest tubes and
replace the sum over the rest of the tubes by an integral.

\begin{figure}
  \centering
  \includegraphics[width=0.6\textwidth]{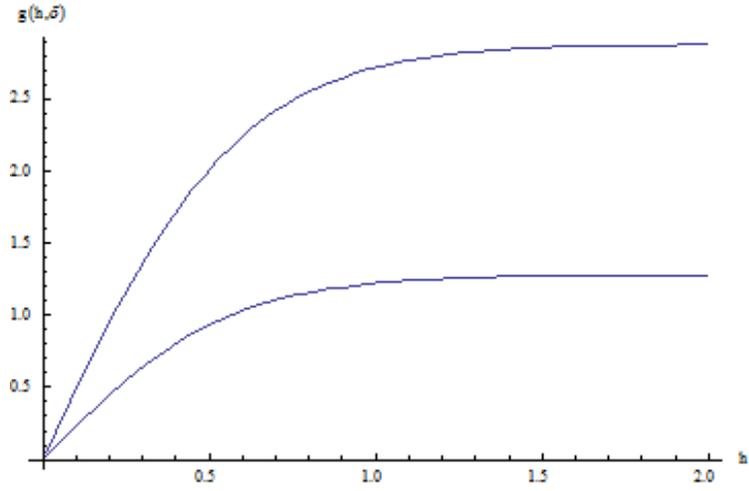}\\
  \caption{Plot of the function $g(h,\delta)$ (i.e. $B_z(\mathbf{x})$ divided by $\frac{\mu_0M_0d}{\pi R}$) at $\delta=3$ (bottom) and $\delta=2$ (up).}\label{17}
\end{figure}

Let us start with the field directly on the $z-$axis. We calculate
separately the contribution of the 7 nearest tubes and integrate out
the contribution of the others. We denote $h=\frac{z-\frac{L}{2}}{R}$
and $\delta=\frac{D}{R}$. The result is:
\begin{equation}
{B_{z}(0,0,\frac{L}{2}+hR)}=\frac{\mu_{0}M_{0}d}{\pi R}g(h,\delta),\label{}
\end{equation}
where \begin{eqnarray}\label{}
g(h,\delta)&=&\frac{\pi}{2}\frac{h}{(1 + h^2)^{\frac{3}{2}}}+ 6\frac{h E(\sqrt{\frac{4\delta}{(\delta+1)^2+h^2}})}{\sqrt{(\delta+1)^2+h^2}[(\delta-1)^2+h^2]} \nonumber\\
&&+2\pi\frac{2}{\sqrt {3}\delta^2} \int_{\sqrt{\frac{7\sqrt3}{2\pi}}\delta}^{\infty}\frac{h E(\sqrt{\frac{4r}{(r+1)^2+h^2}})}{\sqrt{(r+1)^2+h^2}[(r-1)^2+h^2]}rdr.
\end{eqnarray}
The functions $g(h,\delta)$ vs. $h$ at $\delta=3$ and 2 are plotted numerically
in Fig. \ref{17}. They grow from zero asymptotically to $g(\infty,3)=\frac{2\pi^{2}}{9\sqrt{3}}=1.266$ and $g(\infty,3)=\frac{2\pi^{2}}{4\sqrt{3}}=2.849$.
At $h=1$ the fields are already very close to the maxima. The asymptote
corresponds to the uniform field produced by an infinite uniformly-charged
plane. (Note that at distances larger than $L$ the charges at the
plane $z=-\frac{L}{2}$ makes the field zero at infinity.) 

\begin{figure}
  \centering
  \includegraphics[width=0.6\textwidth]{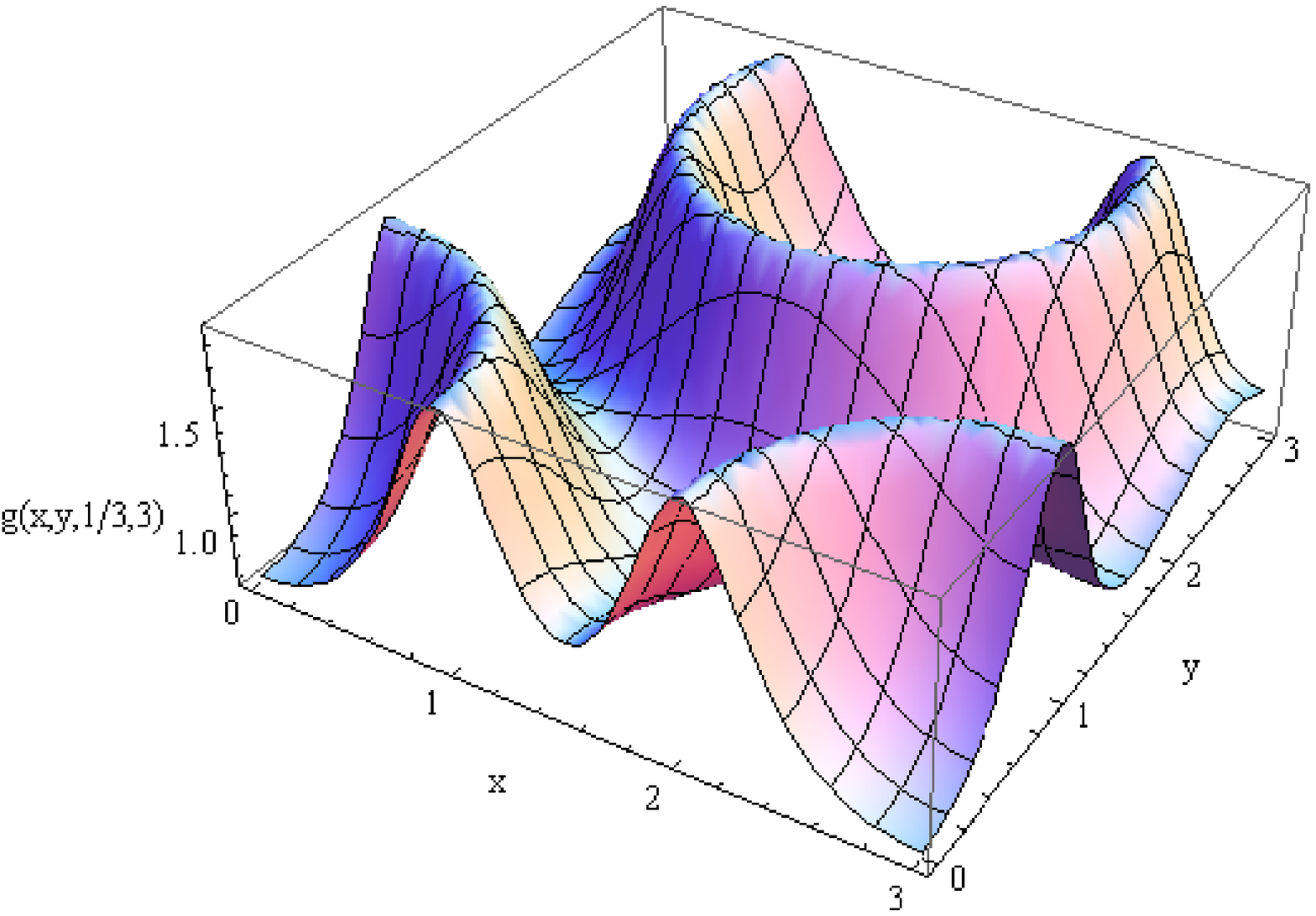}\\
  (a)\\
  \includegraphics[width=0.6\textwidth]{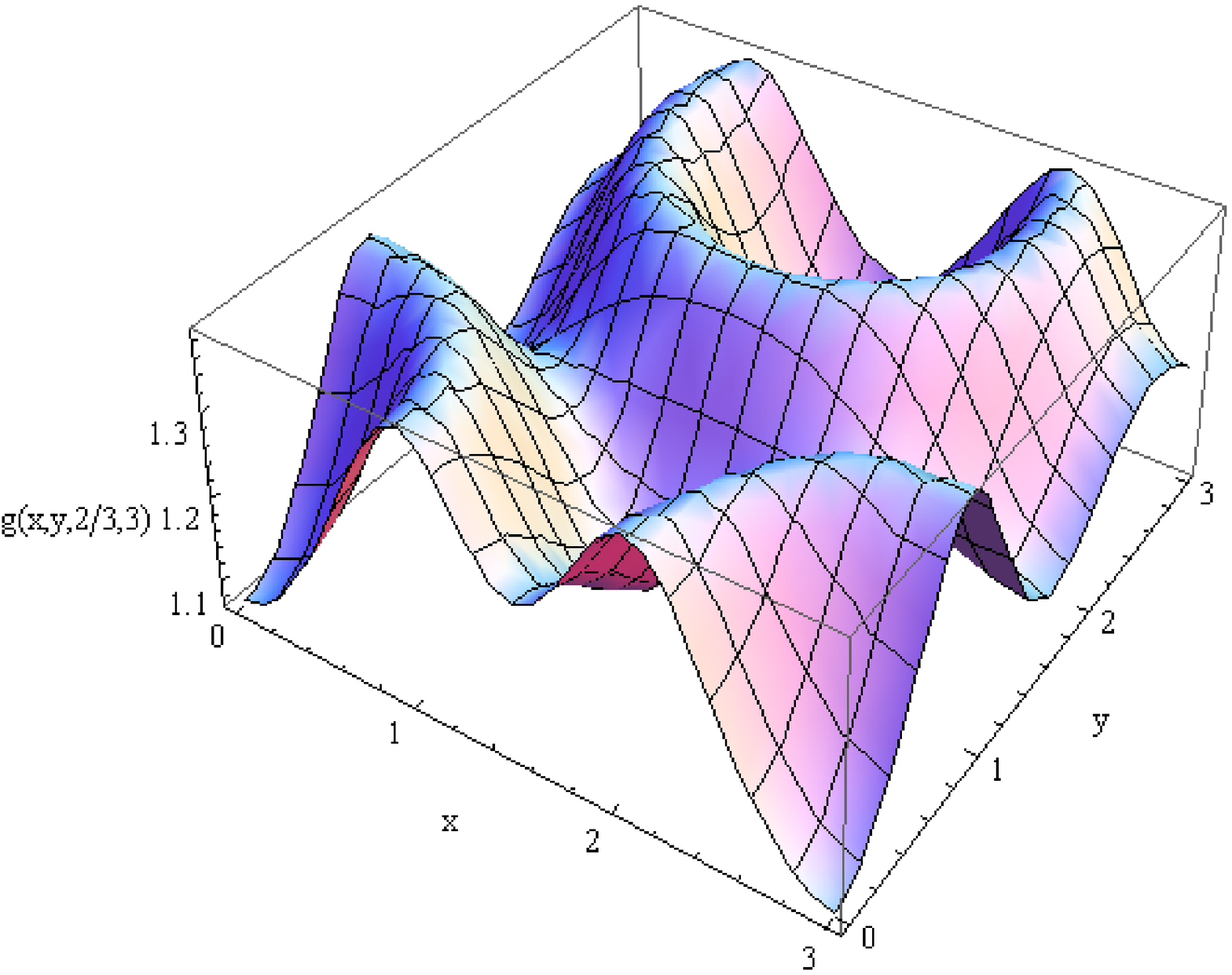}\\
  (b)\\
  \includegraphics[width=0.6\textwidth]{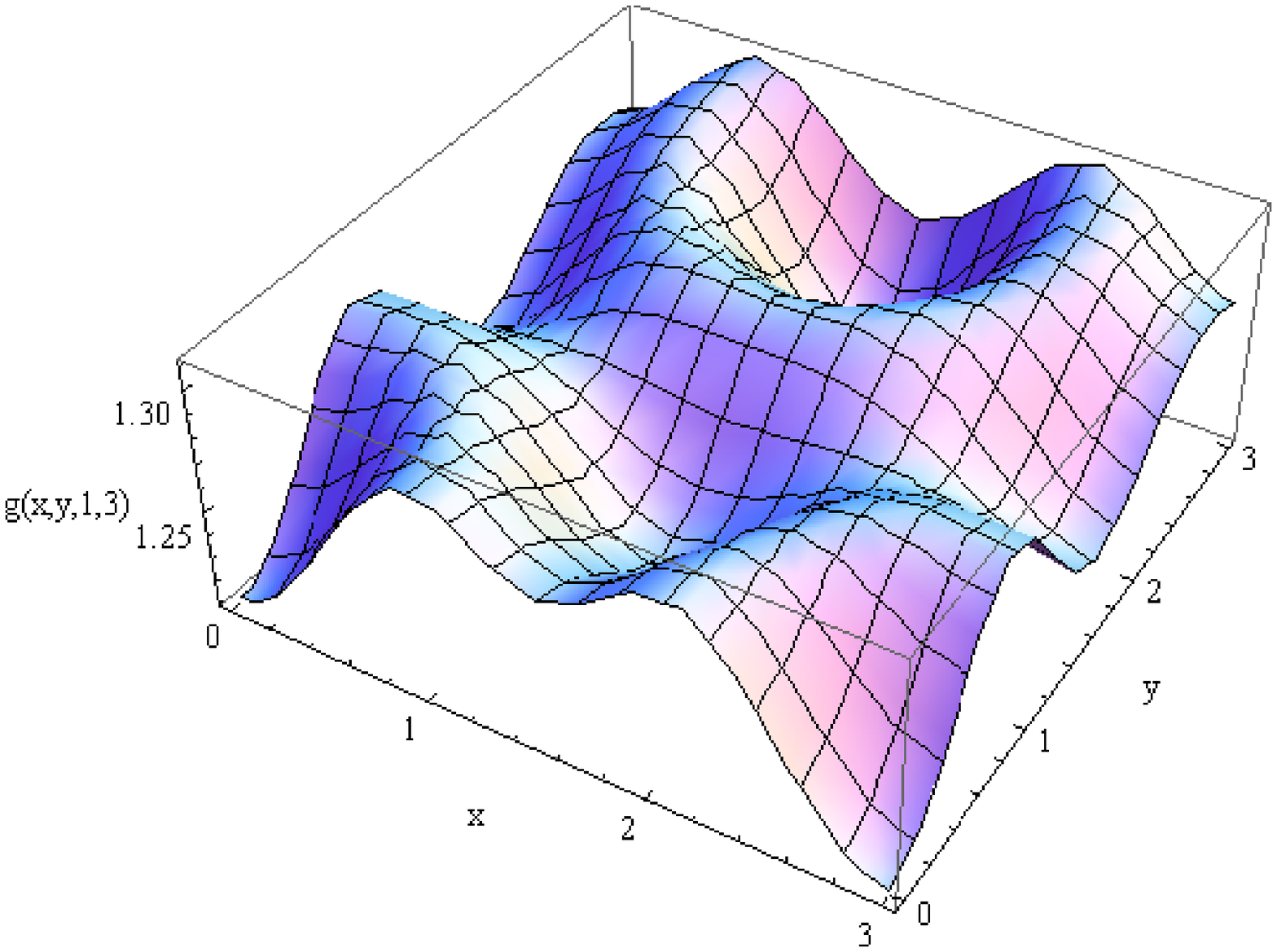}\\
  (c)\\
  \caption{Plot of the function $g_(x,y,h,\delta)$ vs. $r$ at $\delta=3$, and at (a) $h=1/3$ (thick); (b) $h=2/3$ (solid); (c) $h=1$(dashed).}\label{18}
\end{figure}

\begin{figure}
  \centering
  \includegraphics[width=0.6\textwidth]{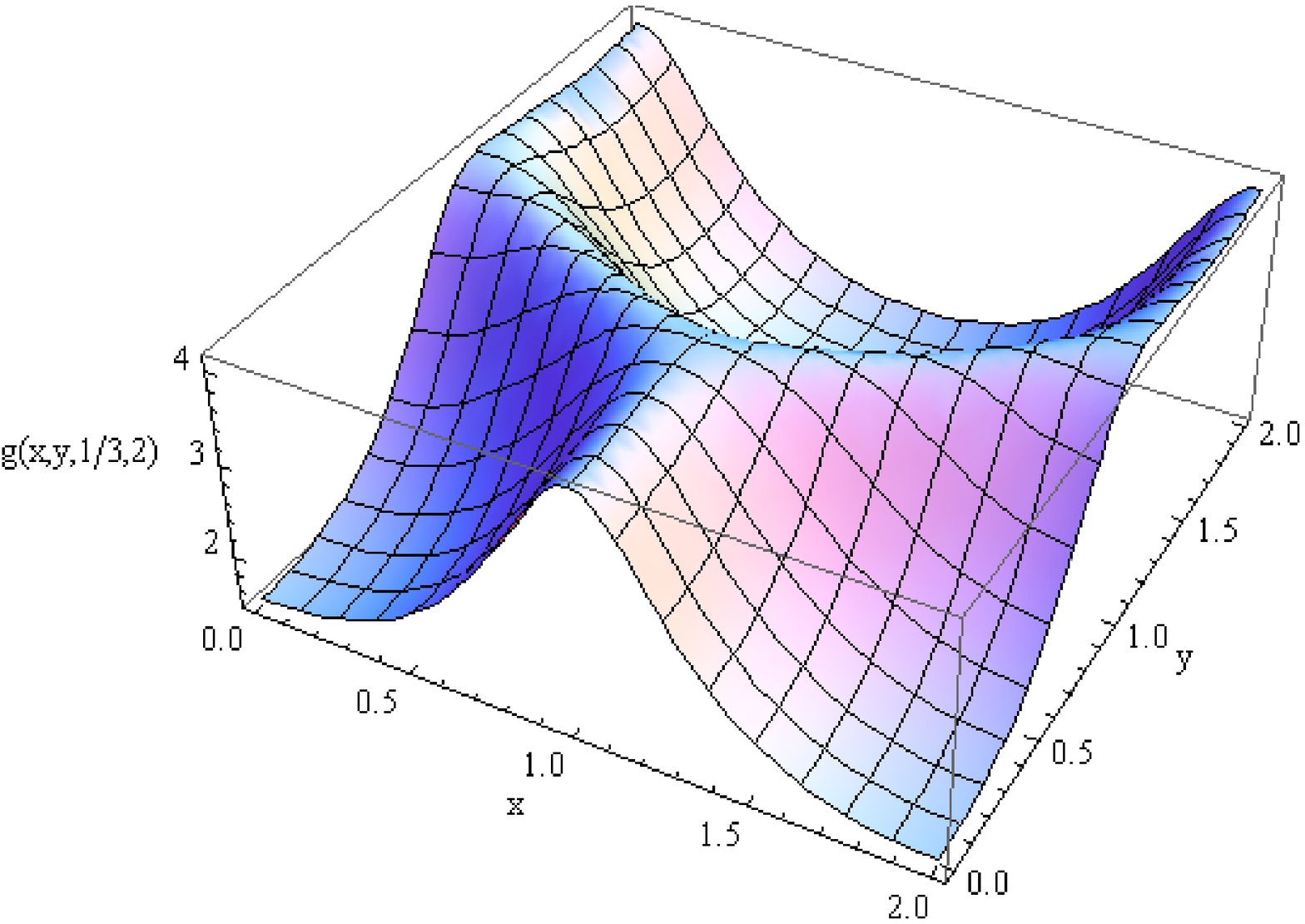}\\
  (a)\\
  \includegraphics[width=0.6\textwidth]{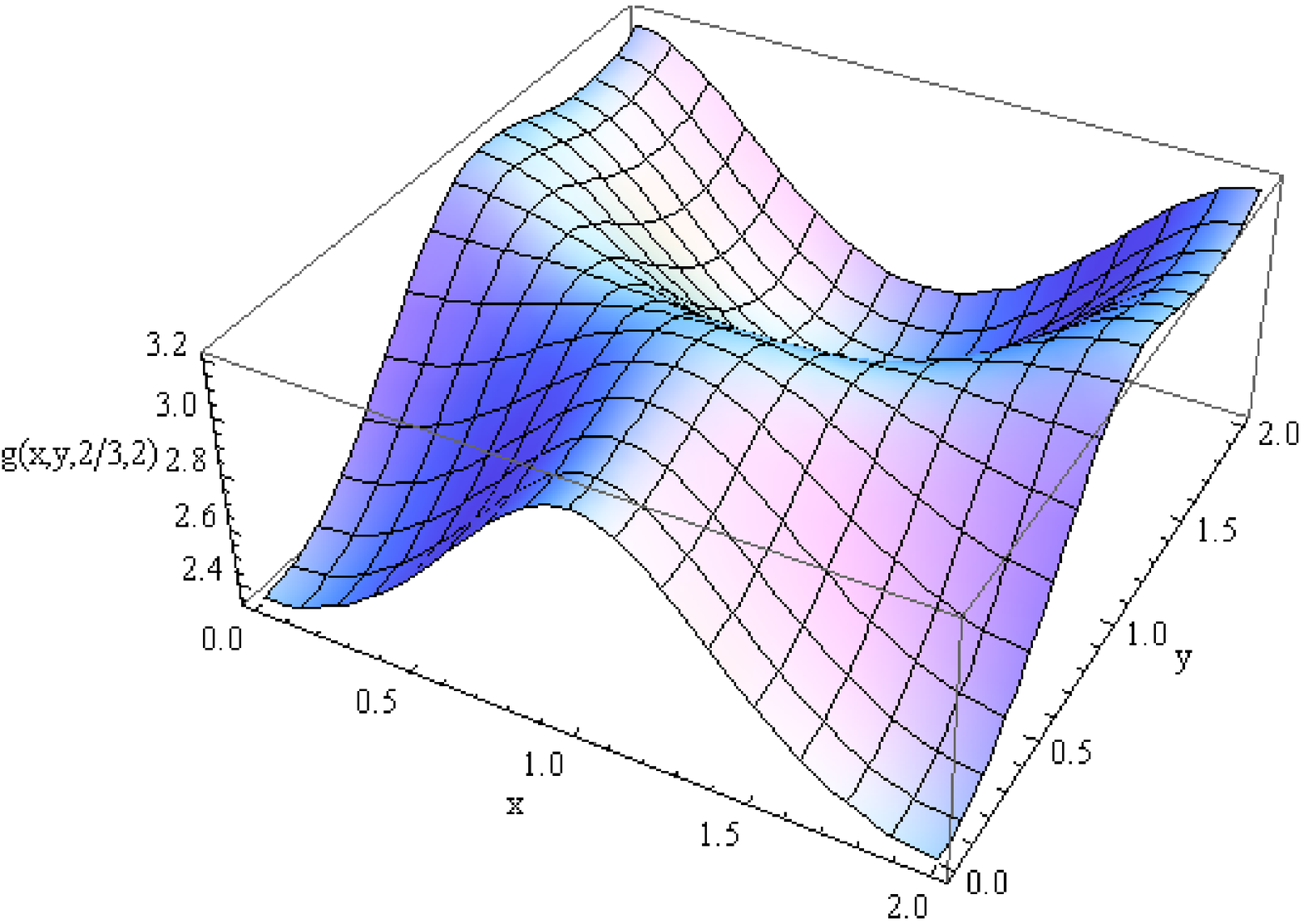}\\
  (b)\\
  \includegraphics[width=0.6\textwidth]{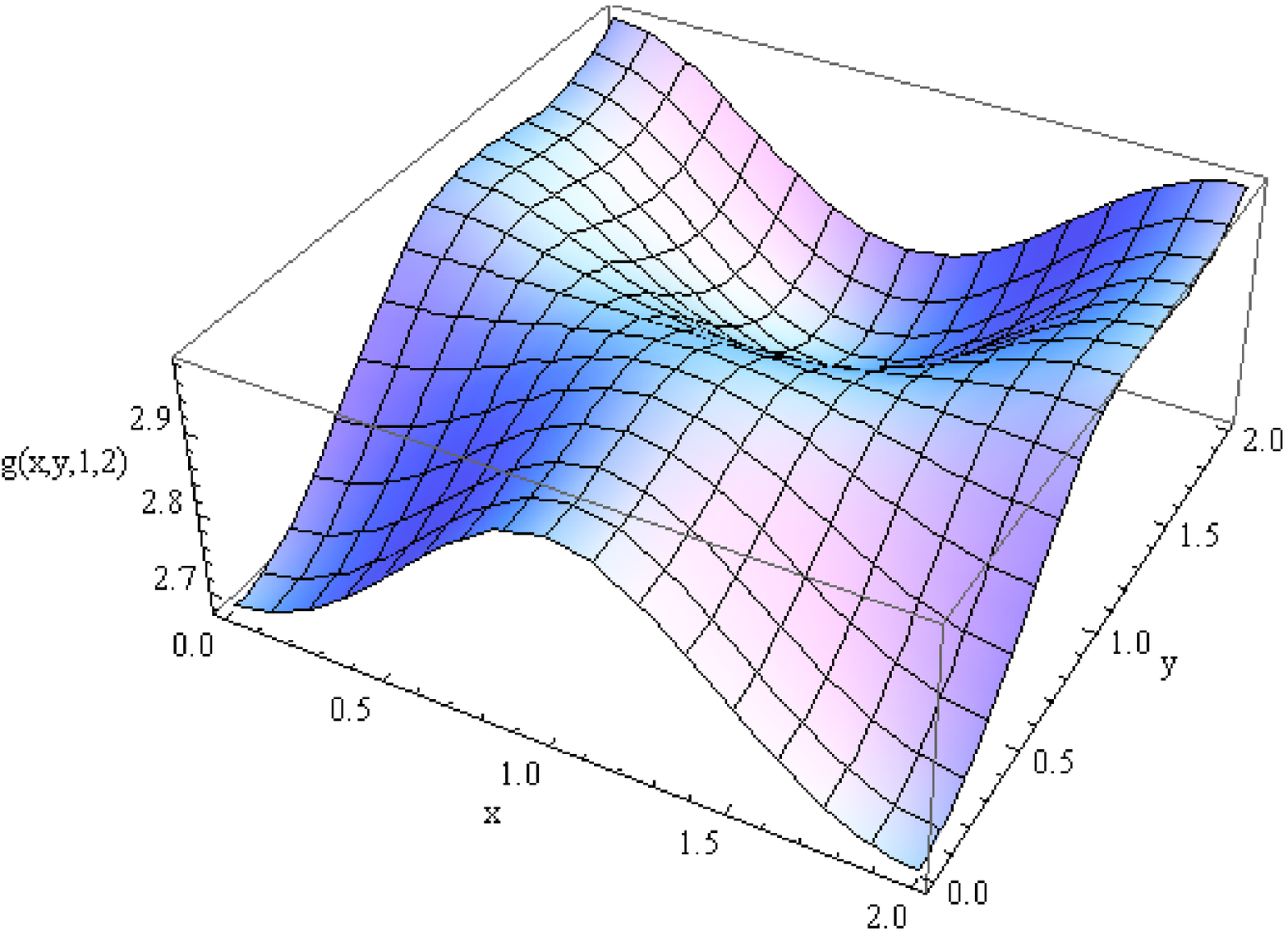}\\
  (c)\\
  \caption{Same as Fig. \ref{18}, but at $\delta=2$.}\label{19}
\end{figure}

For the field other not necessary on the $z-$axis, we calculate separately the contribution of the 12 nearest tubes centered at $(x,y)=(D/2,\sqrt 3 D/2)$, and integrate out the rest. Figs. \ref{18} and \ref{19} are the 3-dim graphs for the function $g(x,y,h,\delta)$ ($B_z$ divided by $\frac{\mu_{0}M_{0}d}{\pi R}$) in the region $0<x<D, 0<y<D$ for $\delta=3$ and 2 (dense filling), respectively. Both are plotted for $h=1/3,2/3,1$. The ridges lie above the tube edges, as expected. The $\delta=2$ cases in general have stronger fields than the corresponding $\delta=3$ cases. Comparing with Fig. \ref{16}(a) we see that the field produced by an array of tubes has a larger strength and varies slower than the field of a single tube.

Again the EDW change the results smearing the field singularities.
Another source of deviations from the simplified model considered
above is the dipolar interaction between tubes. It tends to establish
magnetization in each nearest pair of the tubes oppositely. However,
a rather small magnetic field is sufficient to stabilize the parallel
orientation of magnetization in the tubes.


\section{CONCLUSIONS}

We calculated analytically the exchange and the dipolar energies of
different states for a ferromagnetic long, thin-walled nanotube with
zero crystal anisotropy. In the approximation of infinitely long tubes
at zero external field we have found two possible stable states: the
parallel state and the vortex state. Which of them has lower energy
depends on the dimensionless ratio $\frac{\mu_{0}M_{0}^{2}R^{2}d}{AL}$.
For a long tube with small radius and thickness the parallel state
is favored; increasing the radius and thickness it is possible to
stabilize the vortex state. For a tube in either of these two states,
a domain wall in the middle is proved to be always not energy-favorable.
But for the parallel state there can exist a stable half-DW structures
at the edges of the tube if the radius and thickness are large enough.
In an external field parallel to the tube in the vortex state no hysteresis
appears, whereas the tube in the parallel state subject to the same
field displays a rectangular hysteresis loop. When a field perpendicular
to the tube axis is applied, the vortex state turns into the onion
state at a critical field. This transition is accompanied by a jump
of the magnetization. We also calculated the stray magnetic field
generated by single magnetic tube and the periodic array of the tubes.
It displays a singularity near the edges especially in strong external
magnetic field. The singularities are partly smeared out by the edge
domain walls at external magnetic field decreasing.

\section*{ACKNOWLEDGEMENTS}

We thank Donald G. Naugle, Zhiyuan Wei and Wenhao Wu for discussions of experimental situations and Artem G. Abanov, Fuxiang Li and Peng Zhou
for theoretical discussions. We especially thank Wenhao Wu who led the experimental works of nanotubes that stimulated our study, and we owe our original motivation of this study to him.

\newpage

\phantom{}

\end{document}